\documentclass[twocolumn,tighten,twocolappendix]{aastex701} 

\usepackage{apjfonts} 
\usepackage{physics}
\usepackage{multirow}
\usepackage{graphicx}
\usepackage{amsmath}
\usepackage{amssymb}
\usepackage{color}
\usepackage{mathtools}
\usepackage{ulem} 
\usepackage[T1]{fontenc}
\usepackage{makecell}

\graphicspath{{./}{figures/}}
\newcommand{\revision}[1]{{#1}}

\newcommand{\Myr}{{\rm Myr}}
\newcommand{\Gyr}{{\rm Gyr}}
\newcommand{\pc}{{\rm pc}}
\newcommand{\kpc}{{\rm kpc}}
\newcommand{\Msun}{M_\odot}

\newcommand{\kms}{{\rm km\,s}^{-1}}
\newcommand{\eV}{{\rm eV}}

\begin{document}
\shorttitle{Kinematics of Stellar Streams from Globular Clusters}
\shortauthors{Weatherford \& Bonaca}

\title{Kinematics of Stellar Streams from Globular Clusters Depend on Black Hole Retention and Star Mass:\\A Selection Effect for Dark Matter Inference}
%

\author[0000-0002-9660-9085]{Newlin C. Weatherford}
\affil{Observatories of the Carnegie Institution of Washington, 813 Santa Barbara Street, Pasadena, 91101, CA, USA}
\email{nweatherford@carnegiescience.edu}

\author[0000-0002-7846-9787]{Ana Bonaca}
\affil{Observatories of the Carnegie Institution of Washington, 813 Santa Barbara Street, Pasadena, 91101, CA, USA}
\email{abonaca@carnegiescience.edu}

\begin{abstract}
Velocity dispersion ($\sigma$) in stellar streams from globular clusters (GCs) is sensitive to heating by Galactic substructure, including dark matter (DM) subhalos. Recent studies have compared $\sigma$ in observed and modeled streams to probe DM properties, but have relied on stream models that neglect strong encounters, black holes (BHs), and mass segregation in GCs. Such phenomena may inflate stream $\sigma$ or introduce selection effects---e.g., a $\sigma$ that depends on star mass ($m$). We investigate this prospect using Monte Carlo $N$-body simulations of GCs under static Galactic tides to generate mock streams with realistic mass and velocity distributions. We find $\sigma$ correlates with $m$, especially after core collapse (the GC's observable increase in central density upon ejecting its BHs), rising from $1.2$--$2.2\,{\rm km\,s}^{-1}$ between $m=0.3$--$0.8\,\Msun$, with typical kinematic cuts on stream membership. Similar in magnitude to heating by Galactic substructure, this enhancement occurs because the GC's loss of BHs allows its most-massive stars to occupy its dense center, raising their likelihood of strong ejection via binary interactions and adding broad, exponential wings to the stream's velocity distribution. Streams' kinematics thus probe properties (density, BH retention) of their progenitor GCs. Our results also imply observations of streams from some GCs, especially those not subject to highly episodic mass loss, may select for higher $\sigma$ than predicted by models neglecting $\sigma$'s $m$-dependence. This would cause observed $\sigma$ in streams---already on the low side of expectations for cold DM---to further favor alternatives such as warm or ultralight DM.
\end{abstract}

\section{Introduction} \label{S:intro}
Galactic archeologists have detected abundant Milky Way (MW) substructure, including ${>}120$ stellar streams \citep{Helmi2020, Mateu2023, BonacaPriceWhelan2025}. These strands of comoving stars are debris from dwarf galaxies accreted and tidally disrupted by the MW, and from dissolving globular clusters (GCs) born in those dwarfs (ex situ) or in the MW itself \citep[in situ;][]{Bonaca2021}. 20 streams have been firmly traced to surviving GCs
and many more are likely from fully dissolved GCs \citep{BonacaPriceWhelan2025}.

As kinematically cold structures sensitive to perturbative heating, streams are excellent probes of the MW's mass profile, substructure, and assembly history \citep{Koposov2010, Bonaca2014, Kupper2015, Bovy2016, Bonaca2018}. They also provide clues to the nature of dark matter (DM). Canonical cold DM cosmology predicts clumpy DM halos \citep{Klypin1999, Moore1999} with more numerous and centrally dense (cuspy) subhalos, on average, than warm DM \citep{Bode2001, Lovell2014}, ultralight DM \citep{Hu2000,Hui2017}, or self-interacting DM \citep[which also yields more diverse subhalo densities;][]{SpergelSteinhardt2000, TulinYu2018}. So galaxy models with cold DM predict frequent encounters with subhalos strongly perturb streams, yielding streams with higher width and internal velocity dispersion $\sigma$ \citep{Johnston2002, Ibata2002, Mayer2002, Carlberg2009}---and more numerous and pronounced irregularities, such as gaps \citep{SiegalGaskins2008, Carlberg2011, Yoon2011}---than most alternative DM models.

Since these first studies, most efforts to use streams to probe DM focus on \textit{morphology}---gaps, kinks, and spurs in stream density profiles left by passing subhalos \citep{Carlberg2012a, Carlberg2012b, Carlberg2013a, Carlberg2013b, Carlberg2016a, Ibata2016, Helmi2016, Erkal2016, Bovy2017, Bonaca2019, Carlberg2020, Banik2021, Menker2024, Adams2024, Zhang2025, Nibauer2025a}. Broad surveys with full kinematic detail (especially radial velocities) remain limited, so it is unsurprising that fewer studies explore how such encounters affect stream \textit{kinematics} \citep{Carlberg2015, Carlberg2016b, CarlbergAgler2023, Carlberg2024a, Carlberg2024b, Carlberg2025, Nibauer2025b}. Yet these works show $\sigma$ is a powerful metric for DM inference. Heating by subhalos widens the velocity distribution in GC streams, adding to a dominant cold component ($\sigma \sim 1$--$2\,\kms$) exponential wings whose $\sigma$ is sensitive to the DM model \citep[$6\,\kms$ for cold DM and $3$--$4\,\kms$ for 5.5--7~keV warm DM;][]{Carlberg2024b}. Subhalo heating is thus one of several possible explanations for predominately cold streams enveloped by diffuse warm ``cocoons" \citep[e.g., GD-1;][]{Malhan2019b, Valluri2025, Carlberg2025}.

Stream $\sigma$ is doubly useful for GCs born ex situ, which experience tidal stripping in their parent halo prior to accretion by the MW \citep{Carlberg2018}. $\sigma$ then probes the form of the parent, too; models by \cite{Malhan2021a, Malhan2022b} show GCs born ex situ in cuspy halos produce streams with higher $\sigma$ ($3$--$10\,\kms$, depending on halo mass $M_h$) than GCs born in cored halos ($1$--$4\,\kms$) or in situ (${\lesssim}1\,\kms$). They note many streams from GCs have $\sigma\lesssim 3\,\kms$, ruling out origins in cuspy DM halos with $M_h\gtrsim10^8\,\Msun$. This point \citep[see also][]{Gialluca2021, Grillmair2025} applies to Fj\"{o}rm, Gj\"{o}ll, and Sylgr (from the GCs M68, NGC~3201, and NGC~5024, respectively), GD-1 and Phlegethon (unknown progenitors), and NGC~0288's stream. As these streams' orbits indicate ex-situ origins \citep{Bonaca2021}, their low $\sigma$ is in soft tension with cuspy cold DM, especially given the expectation of further heating by subhalos after accretion.

The low $\sigma$ of many streams and the relatively small (few-$\kms$) impacts of different DM models or parent halos mean that other effects capable of shifting $\sigma$ by ${\gtrsim}0.5\,\kms$ are relevant when using $\sigma$ to infer the properties of DM or parent halos. These include heating by baryonic MW substructures, spurious enhancement of $\sigma$ from unresolved binaries or star properties (see Section~\ref{S:discussion}), and internal GC dynamics. 

Myriad dynamics drive escape from GCs \citep[for a thorough review, see][]{Weatherford2023}, but most stars reach the requisite energy via two-body relaxation \citep[the slow random-walk in energy from many weak encounters with other stars;][]{BinneyTremaine2008}, or via strengthening external tides during perigalacticon \citep{BaumgardtMakino2003} or passage by/through some other perturber \citep[e.g., the Galactic disk;][]{Ostriker1972}. Each passage temporarily shrinks the GC's tidal boundary, lowering the threshold energy for escape, and tidally shocks stars with orbital periods longer than the passage. This mostly strips the GC's outermost stars, but the lower escape energy aids ejection from its core, too. Also, since relaxation is fastest in the core, most stars in GCs not subject to highly fluctuating tides first obtain enough energy to escape while passing through it \citep{SpitzerShapiro1972, Weatherford2023}. Once unbound, they take about a relaxation time (${\sim}100\,\Myr$) to escape \citep{Weatherford2024}, bouncing about inside the tidal equipotential surfaces enclosing the GC before exiting via openings centered on the L1/L2 Lagrange points.
Since relaxation and gently fluctuating tides only weakly unbind stars, these openings are narrow for most ejecta, tightly collimating them into twin tidal tails that together form a stream with $\sigma\lesssim2\,\kms$.


\revision{Relaxation is not the only phenomenon in GCs relevant to streams, but it is dominant. So even the few prior studies of stream kinematics that explicitly model the GC's internal dynamics with direct $N$-body codes \citep{Malhan2021a, Malhan2022b, Carlberg2024a, Carlberg2024b, Carlberg2025} hasten computation by using softening lengths 1--2$\,\pc$ and assuming all stars have equal mass.}
They thus neglect some \revision{key secondary} drivers of GC evolution: strong encounters (which occur on scales ${\ll}1\,\pc$), mass segregation, and stellar-mass black holes (BHs). While most impactful in the core, these factors may significantly affect some streams. Strong encounters in the core can eject up to ${\sim}10\%$ of escapers from dense GCs, their higher energies leading to less-collimated escape and a wider, hotter subcomponent in the stream \citep{Weatherford2024}. And the BHs that form in a GC are essential heat sources supporting the core against collapse \citep[e.g.,][]{Merritt2004b, Mackey2007, Mackey2008, BreenHeggie2013, Kremer2019a, CMCCatalog}. After mass-segregating to the center, they pair up into binaries that undergo strong encounters with other BHs and stars, hardening the binaries and accelerating the bodies involved. These encounters eject BHs from the core; those that do not entirely escape the GC resegregate back to its center, passing more energy to surrounding stars. This greatly enhances stream density if it expands the GC's half-mass radius too close to its tidal boundary, potentially explaining the Pal~5 stream's high density \citep{Gieles2021, Gieles2023, Roberts2024}.

Selection effects may also arise from these dynamics. In particular, $\sigma$ in streams may depend on star mass $m$, as in GCs. 
\revision{While the \cite{Spitzer1969} instability prevents full energy equipartition \citep{Trenti2013, Bianchini2016}, many observed GCs exhibit modest $m$--$\sigma$ anticorrelations, with $\sigma\propto m^{-\eta}$ and $\eta\approx0.1$--$0.4$ \citep{BaumgardtCatalog2020,Watkins2022,Ziliotto2025}.}
\revision{In concert with mass segregation, near ubiquitous in observed GCs \citep{Dacosta1982, Goldsbury2013, BaumgardtCatalog2020, Weatherford2020}, this means a GC's}
lowest-$m$ stars tend to orbit the most energetically with the furthest apocenters, easing escape. So low-$m$ stars escape earliest and model streams exhibit slight $m$ gradients, especially close to dissolution of the highly mass-segregated core \citep{Webb2022}. It is thus tempting to imagine a star's final ejection kick, when stacked with its $m$-dependent orbital energy, would cause $\sigma$ in the stream to anticorrelate with $m$. But the notion of a single discrete kick only applies to strong encounters; most ejecta (from relaxation and tides) barely drift across the tidal boundary, regardless of $m$, leaving little prospect of an $m$--$\sigma$ anticorrelation in the stream. $\sigma$ may instead \textit{correlate} with $m$, as the highest-$m$ stars, having sunk deepest into the core, would be ejected the most often by the strong encounters there. As we shall show later, this possibility turns out to be correct.

Stream observations are often highly magnitude-limited, so any dependence of $\sigma$ on $m$, and thereby luminosity, may bias the observationally inferred $\sigma$---and any DM inference based on it. This would be especially relevant to the distant, low-eccentricity streams furthest from the Galactic center---the very streams best suited to probing the MW's own DM halo \cite[][]{Sandford2017}---and to extragalactic streams, essential to probing DM halos more broadly \citep{Pearson2019}. This motivates investigation of any potential $m$ dependence in the kinematics of streams from dense star clusters. To do so we simulate GCs and their streams with all relevant collisional dynamics, stellar evolution, and a realistic stellar mass function and stream velocity distribution.

\begin{figure*}[ht!]
\centering
\includegraphics[width=\linewidth]{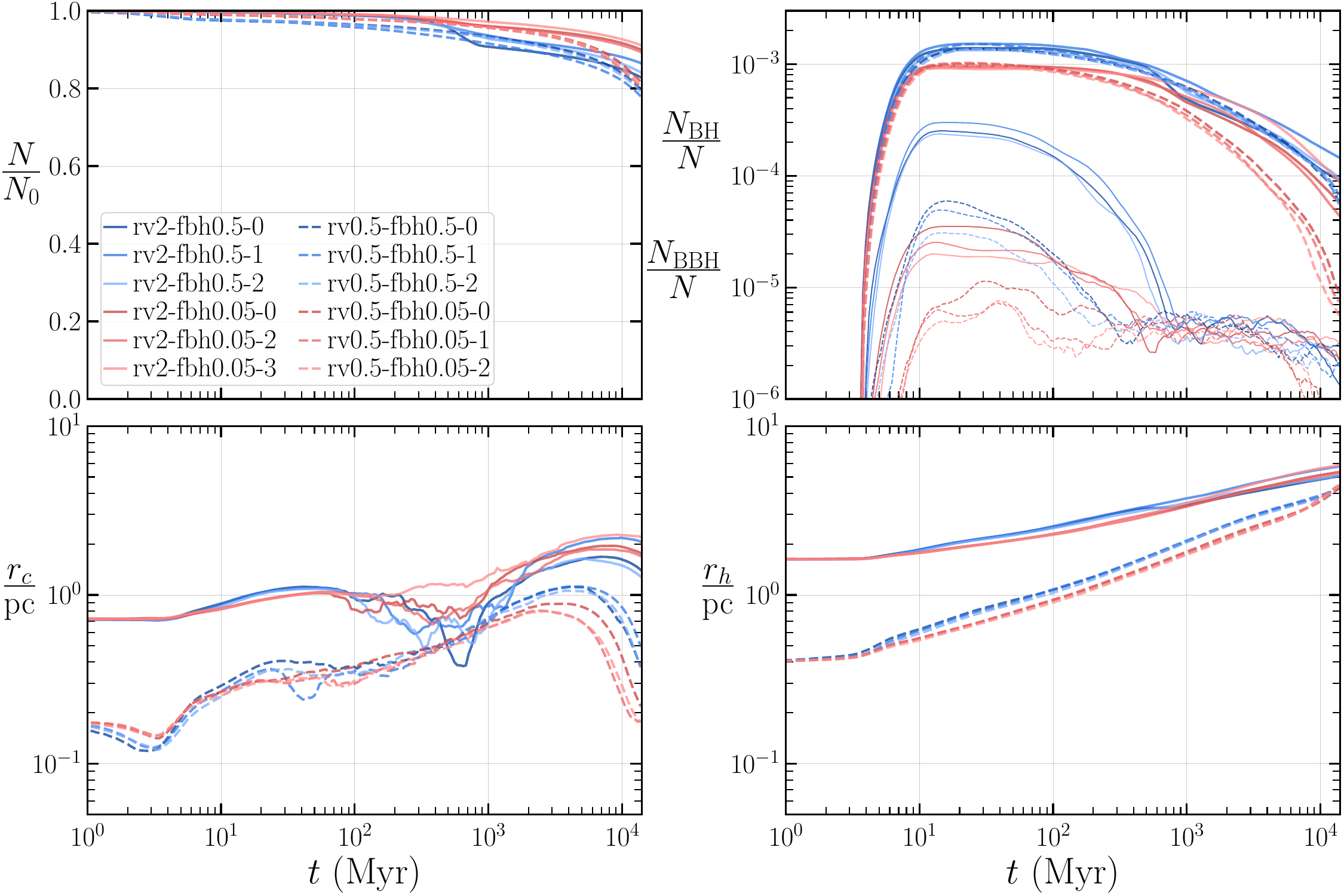}
\caption{Rolling average time evolution of GC properties in our simulations. Line style (dashed, solid) distinguishes simulations with different initial $r_v=(0.5,2)\,\pc$ and coloration (red, blue) indicates the initial high-mass binary fraction $f_{b,h}=(5\%,50\%)$, with specific shading distinguishing independent statistical realizations of the same $(r_v,f_{b,h})$ combination. \textit{Top left}: number of particles (single stars plus binaries) relative to their initial number. \textit{Top right}: number of BHs and BH--BH binaries (BBHs). \textit{Bottom left}: density-weighted core radius $r_c$ \citep{CasertanoHut1985}. 
\textit{Bottom right}: half-mass radius $r_h$. The BHs kinematically heat the GC via strong binary interactions and repeated segregation to the GC's center, inflating $r_h$ and supporting $r_c$ against collapse. The permanent drop in $r_c$ starting at $t\approx5\,\Gyr$ for GCs born with $r_v=0.5\,\pc$ (dashed curves) reflects the transition to an observably core-collapsed state that occurs once the GC has ejected the vast majority of its BHs.}
\label{fig:macro_evolution}
\end{figure*}

The paper is organized as follows. We describe our GC simulations and mock stream generation in Section~\ref{S:methods}. We analyze the stream velocity distributions in Sections~\ref{S:results_velocity_distribution}--\ref{S:results_curve_fitting}, including contributions of specific escape mechanisms and the dependence on $m$, and GC density and age (tied to BH retention and mass segregation). We demonstrate an $m$--$\sigma$ correlation in our streams in Section~\ref{S:results_velocity_dispersion} and discuss our results versus other factors affecting $\sigma$ (Section~\ref{S:discussion_other_factors}), the $\sigma$ observed in GC streams (Section~\ref{S:discussion_observed_dispersions}), and direct $N$-body simulations (Section~\ref{S:discussion_nbody}). We also comment on binaries in our streams (Section~\ref{S:discussion_binaries}) before summarizing our findings in Section~\ref{S:summary}.

\section{Numerical Models} \label{S:methods}
\subsection{Globular Cluster Simulations} \label{S:gc_simulations}
To reliably investigate the prospect of any $m$ dependence in a GC stream's velocity distribution, we must fully simulate the GC's internal dynamics, including strong encounters that eject stars at high speeds. We do so using our publicly available \texttt{Cluster Monte Carlo} code \citep[\texttt{CMC};][]{CMCRelease}. Developed over 25 years, \texttt{CMC} is a state-of-the-art implementation of the orbit-averaged Monte Carlo method \citep{Henon1971a,Henon1971b} and can simulate typical MWGCs in ${\sim}10^3$--$10^4$ times less computing time than a direct $N$-body approach while still accurately capturing the internal dynamics relevant to escape. This includes the evaporative (low-speed) escape mechanisms most relevant to stream formation: two-body relaxation \citep{Joshi2000,Pattabiraman2013} coupled with the Galactic tide \citep{Joshi2001, Chatterjee2010}. \texttt{CMC} also features many high-speed ejection mechanisms, such as strong binary-mediated interactions, three-body binary formation \citep[from three singles;][]{Morscher2013, Morscher2015}, gravitational-wave-driven merger \revision{kicks} \citep{Rodriguez2018a,Rodriguez2018b}, and supernova kicks during stellar evolution \citep[with \texttt{COSMIC};][]{Breivik2020}. We focus on the the first \revision{two} of these mechanisms---the most relevant to \revision{observable high-speed stellar ejecta in streams from old} GCs---but for more thorough descriptions of \texttt{CMC} and its dynamics relevant to escape, see \cite{CMCRelease} and \cite{Weatherford2023}, respectively. While \texttt{CMC} allows physical collisions, we do not consider asymmetric mass loss in such events that may eject the remnant at high speed. Nor do we consider ejection by a central intermediate-mass BH, which may be present in some GCs.

For simplicity and consistency with earlier models by \cite{CMCCatalog} and \cite{Weatherford2023, Weatherford2024}, we run a grid of twelve GC simulations to an age of $14\,\Gyr$---three statistically independent realizations each of four models designed to represent archetypal core-collapsed and non-core-collapsed GCs (explained shortly). These models differ only in their initial virial radius, $r_v=(0.5,2)\,\pc$, and binary fraction for massive stars, $f_{b,h}=f_b(m>15\,\Msun)=(5\%,50\%)$; the binary fraction at lower masses is a flat $5\%$, typical of MWGCs at present \cite[][]{Milone2012}. We chose to vary $f_{b,h}$ thinking $f_{b,h}=50\%$---more in accord with observations of young star clusters \citep{Sana2009, Sana2012, MoeDiStefano2017}---may enhance strong ejections. Yet there turned out to be no significant enhancement at the ages typical of MWGCs.

The more important parameter is $r_v$. The higher initial density in the simulations with $r_v=0.5\,\pc$ leads to faster dynamics; by an age corresponding to the present (${\approx}12\,\Gyr$), they eject almost all of their central BH population. The loss of heating from BH dynamics allows the GC's core to contract, forming a cuspy inner surface density profile characteristic of an observably \textit{core-collapsed} state \citep[${\approx}20\%$ of surviving MWGCs;][]{Trager1995}. But the simulations born with $r_v=2\,\pc$ are initially puffy enough to retain a substantial fraction of their BH populations all the way to the present, preserving a flat inner surface brightness profile characteristic of a \textit{non-core-collapsed} state (${\approx}80\%$ of MWGCs).

All other initial conditions are chosen to be roughly average among MWGCs \citep{CMCCatalog}. Specifically, they have an initial total number of particles (single stars plus binaries) $N_0=8\times10^5$, Galactocentric distance $R_g=8\,\kpc$, and metallicity $Z=0.1\,Z_\odot$. Initial positions and velocities of stars draw from a \cite{King1962} profile with concentration $W_0=5$. Stellar masses (primary mass, in the case of a binary) draw from the canonical \cite{Kroupa2001} stellar initial mass function from $0.08$--$150\,\Msun$ while secondary masses, assigned randomly to $N_0\times f_b$ stars, draw uniformly in mass ratio from $q\in[0.08,1]$ \citep{DuquennoyMayor1991}. Binary orbital periods draw from a distribution flat in log scale from near contact to the hard/soft boundary \citep{DuquennoyMayor1991} and eccentricities are thermal \citep{Heggie1975}. We assume all neutron stars born in core-collapse (electron-capture) supernovae receive natal kicks drawn from a Maxwellian with dispersion $\sigma=265\,{\rm km\,s}^{-1}$ ($20\,{\rm km\,s}^{-1}$), ejecting most of them from the GC at birth. Natal kicks for BHs share the neutron stars' core-collapse kick distribution, but reduced in magnitude by the fraction of the stellar envelope's mass that falls back onto the BH via the prescription from \cite{Hobbs2005}---see also \cite{Fryer2012}. This ejects ${\approx}1/2$ of BHs at birth. For further modeling details, see the descriptions by \cite{Weatherford2024} or references therein.

Figure~\ref{fig:macro_evolution} shows how key macroscopic properties of each GC simulation evolve with age, including the number of particles $N$ normalized by the initial value $N_0$ (top left panel), the number of BHs $N_{\rm BH}$ and binary BHs $N_{\rm BBH}$ (top right), the core radius $r_c$ \citep[bottom left, using the density-weighted definition of][]{CasertanoHut1985}, and the half-mass radius $r_h$ (bottom right). Note all curves are rolling averages, smoothing out, e.g., gravothermal oscillations in $r_c$. Line style (dashed, solid) distinguishes simulations with different initial $r_v=(0.5,2)\,\pc$ while coloration (red, blue) indicates initial $f_{b,h}=(5\%,50\%)$, with shading distinguishing independent statistical realizations of the same $(r_v,f_{b,h})$ combination. As intended, the denser models ($r_v=0.5\,\pc$) all contract to an observably core-collapsed state by ${\approx}9$--$12\,\Gyr$, after ejecting most of their BHs. The GCs born with $f_{b,h}=50\%$ also form and initially retain ${\sim}50\%$ more BHs---and 10 times as many binary BHs---than the GCs born with $f_{b,h}=5\%$. This is due to how \texttt{CMC} uniformly samples the binary mass ratio, causing most BH progenitors to receive a secondary that is also massive enough to become a BH. The initial boost to the number of binary BHs disappears after ${\sim}1\,\Gyr$ due to binary destruction via strong encounters and BH mergers, but the additional heating from BH dynamics early in the GC's life somewhat delays the onset of observable core collapse compared to the GCs born with lower $f_{b,h}$. It turns out these differences are small enough that the stream morphology and kinematics from our GCs do not depend much on $f_{b,h}$, so we stack snapshots from all simulations with the same $r_v$ to decrease statistical noise in much of our later analysis.

\begin{figure}[t!]
\centering
\includegraphics[width=\columnwidth]{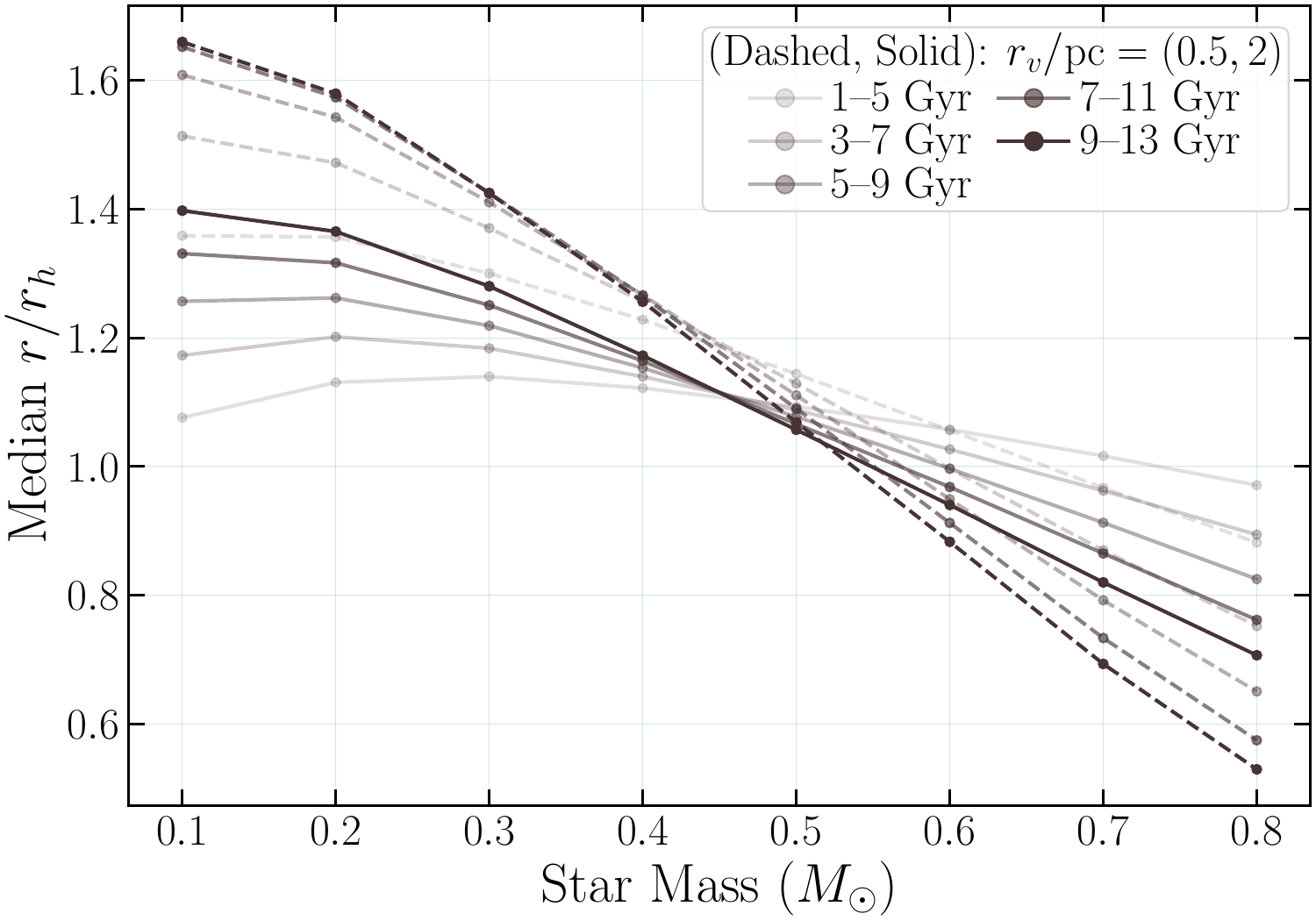}
\caption{Mass segregation in our GCs, measured by stars' median radial positions $r$ versus $m$, excluding compact objects and binaries. Median $r$ is measured in nonoverlapping mass bins (width $0.1\,\Msun$) centered on each data point and is normalized by the GC's global half-mass radius $r_h$. For simplicity, we combine data from all six simulations for each choice of $r_v$: $0.5\,\pc$ (dashed curves) and $2\,\pc$ (solid). Shading distinguishes profiles averaged over distinct GC age ranges (see legend), with darker shades indicating older GCs.}
\label{fig:mass_segregation}
\end{figure}

We characterize mass segregation in our GCs over time in Figure~\ref{fig:mass_segregation}, measured by stars' median radial positions $r$ versus $m$, excluding compact objects and binaries. We combine the data from all six simulations for each choice of $r_v$: $0.5\,\pc$ (dashed curves) and $2\,\pc$ (solid). Shading distinguishes several GC epochs, with darker shades indicating older GCs. The GCs born with $r_v=0.5\,\pc$ are all core-collapsed in the oldest age interval ($9$--$13\,\Gyr$), making it easy to see how GCs post-collapse are significantly more mass-segregated than less dynamically evolved GCs. At this age, for $r_v/\pc=(0.5,2)$, stars in the highest-$m$ bin ($0.75$--$0.85\,\Msun$) have median $r/r_h\approx(0.53,0.71)$ and median $r/\pc\approx(2.3,3.8)$, respectively. As expected classically \citep{HeggieHut2003,BinneyTremaine2008}, mass segregation grows more extreme with increasing dynamical age---i.e., with both true age and with decreasing $r_v$, since denser GCs relax faster.

\begin{figure*}[ht!]
\centering
\includegraphics[width=0.974\linewidth]{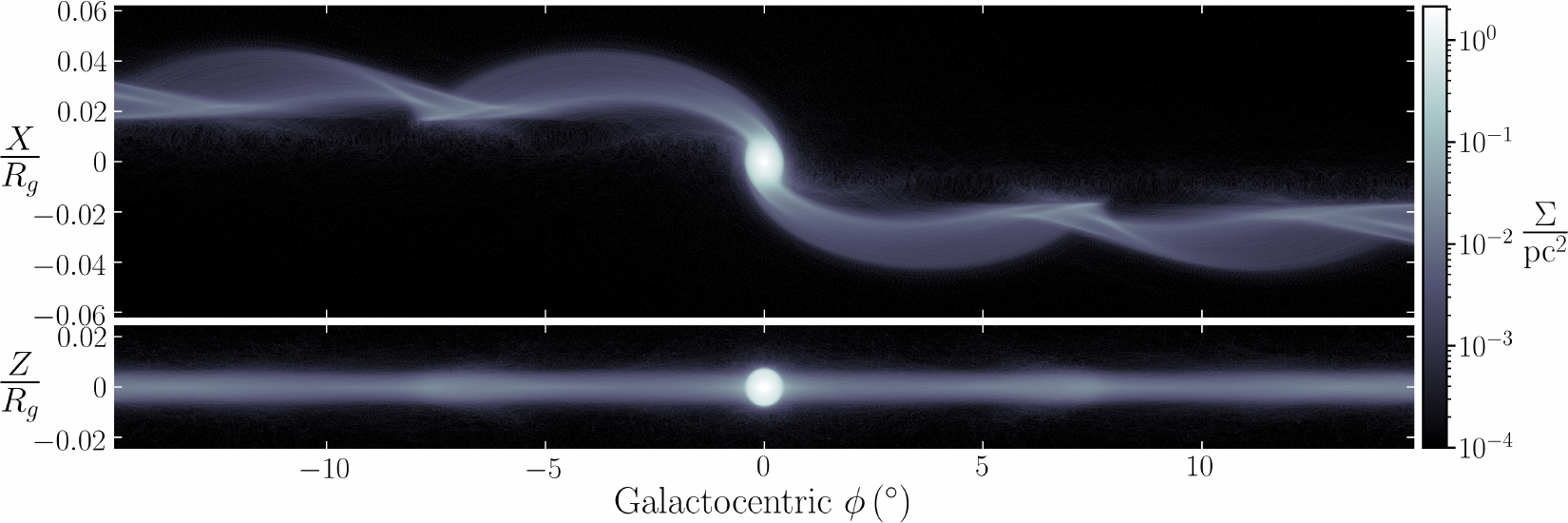}
\caption{Map of the surface number density $\Sigma$ of unbound stars (including those still on their way out of the GC) from one of our GCs born with $r_v=0.5\,\pc$, $f_{b,h}=50\%$. $\Sigma$ is averaged over ages $9$--$13\,\Gyr$ by stacking $401$ snapshots (one every $10\,\Myr$), finely binning by location, and dividing each bin count by $401$ and the bin area in ${\rm pc}^2$. We use Galactocentric cylindrical coordinates $(R,\phi,Z)$, with the radial coordinate expressed as $X=R-R_g$ so that $(X,\phi,Z)=(0,0,0)$ is the GC's center. The top panel is thus a view face-on to the GC's circular orbit after mapping it to the line segment $(X,Z)=0$, $\phi\in[-180^\circ,180^\circ]$. The bottom panel is a panoramic view edge-on to the GC's orbit from the perspective of the Galactic center. \textit{Unless noted otherwise, we select as stream members stars outside the GC's tidal radius $r_t\sim 0.01R_g$ with $\lvert X/R_g \rvert\leq0.05$, $\lvert Z/R_g\rvert\leq0.02$, and $\vert\phi\rvert\leq30^\circ$}. Panel aspect ratios are scaled to avoid stretching the stream; i.e., each pixel has equal side lengths in parsecs.}
\label{fig:density_map}
\end{figure*}

\subsection{Mock Stellar Streams} \label{S:mock_streams}
We follow the methods of \cite{Weatherford2024} to model stellar streams from the \texttt{CMC} simulations, so consult that work for further detail. In short, \texttt{CMC} removes bodies from the simulation the moment they acquire enough energy to \textit{potentially} escape\footnote{In fact, \texttt{CMC} removes a body one it has a bit \textit{more} energy than needed to escape. Developed by \cite{Giersz2008} and tuned to best match escape rates of direct $N$-body models, this feature roughly accounts for ongoing two-body scattering on the way out of the GC. Such encounters make escape slightly harder by scattering some unbound bodies back onto bound orbits \citep{King1959,Baumgardt2001}, so those that \textit{do} escape are preferentially the ones scattered to even higher energy. But this nuance has little impact on the stream \citep{Weatherford2024}.}---usually from within the GC's core, where relaxation is fastest. Since \texttt{CMC}'s dynamics are spherically symmetric, we then randomly and isotropically project the body's position (strictly radial) and velocity (radial and tangential components) into full six-dimensional phase-space coordinates. We then use the Galactic dynamics code \textit{Gala} \citep{GalaCodeRelease} to evolve the trajectories of these escaping bodies in the rotating frame of the combined effective potential of the GC and MW. Since rigorous testing of \texttt{CMC} for eccentric GC orbits is still in progress, we assume the GC circularly orbits the Galactic center with orbital speed $v_c=220\,{\rm km\, s}^{-1}$ \citep[e.g.,][]{BinneyTremaine2008}. For consistency with \texttt{CMC}'s default escape criterion, we fix the MW potential in \textit{Gala}'s trajectory integration to be static, spherical, and logarithmic---$\phi_g(R)=v_c^2\ln(R)$---while the GC potential is allowed to evolve, updated every megayear from analytic fits to the GC potential in each \texttt{CMC} snapshot. As the GC loses mass $M_c$, its tidal radius $r_t$ slowly shrinks with time $t$:
\begin{equation} \label{Eq:tidal_radius}
r_t(t) = \left[\frac{GM_c(t)R_g^2}{2v_c^2}\right]^{1/3}.
\end{equation}

Conceptually, the above procedure reduces to shutting off collisional dynamics for any body in the GC once it becomes unbound and likely to escape (cross beyond $r_t$) within about the GC's half-mass relaxation time, or ${\sim}100\,\Myr$. This decoupling from \texttt{CMC}'s spherically symmetric collisional dynamics is necessary to allow the truly asymmetric tidal field to channel escapers into realistic stellar streams.

To show how we select stream members, Figure~\ref{fig:density_map} maps the surface density $\Sigma$ of unbound stars from one of our GCs born with $r_v=0.5\,\pc$ and $f_{b,h}=50\%$, time-averaged over $9$--$13\,\Gyr$ (see figure caption). The map includes all unbound stars, even those still on their way out of the GC, and uses Galactocentric cylindrical coordinates $(R,\phi,Z)$, with the radial coordinate expressed as $X=R-R_g$ so that $(X,\phi,Z)=(0,0,0)$ is the GC's center. The top panel is a view face-on to the GC's circular orbit after mapping it to the line segment $(X,Z)=0$, $\phi\in[-180^\circ,180^\circ]$. The bottom panel is a panoramic view edge-on to the GC orbit from the Galactic center's perspective. As discussed by \cite{Weatherford2024}, the stream morphology closely matches theoretical expectations; e.g., the width of each tidal tail (${\approx}2.3r_t$) and the low-density channel between them ($2\sqrt{3}r_t$) match analytic predictions and direct $N$-body models \citep{Kupper2008, Kupper2010, Just2009}. Epicyclic overdensities also appear at $\lvert \phi \rvert\approx (7^\circ,14^\circ)$, spaced ${\approx}12.2r_t$ apart---between the estimates by \cite{Just2009} and \cite{Kupper2010} for our choice of $\phi_g$.

We define stream members to be stars outside the tidal radius with $\lvert X/R_g\rvert \leq 0.05$ and $\lvert Z/R_g \rvert \leq 0.02$\revision{, with the limit in each case corresponding roughly to where the stream surface density drops to ${\lesssim}10^{-3}$ of its peak value (i.e., to ${\lesssim}10^{-4}\,\pc^{-2}$. In parsecs, the $X$ criterion ($\lvert X \rvert\leq400\,\pc)$ is very nearly the half-width of the search box for the GD-1 cocoon, and about double the cocoon's half-width half-max \citep{Valluri2025}. The $X$ and $Z$ criteria are thus permissive enough to ensure high completeness while still excluding some especially hot ejecta that would be unlikely to be classified as stream members observationally.} Unless otherwise noted, we also restrict our analysis to $\lvert \phi \rvert \leq 30^\circ$. As $60^\circ$ is ${\approx}8.4\,\kpc$ for our $R_g=8\,\kpc$, this criterion corresponds roughly to the typical length of streams with known GC progenitors \citep[Table~C.1 of][]{BonacaPriceWhelan2025}.

\begin{figure*}[p!]
\centering
\includegraphics[width=0.96\linewidth]{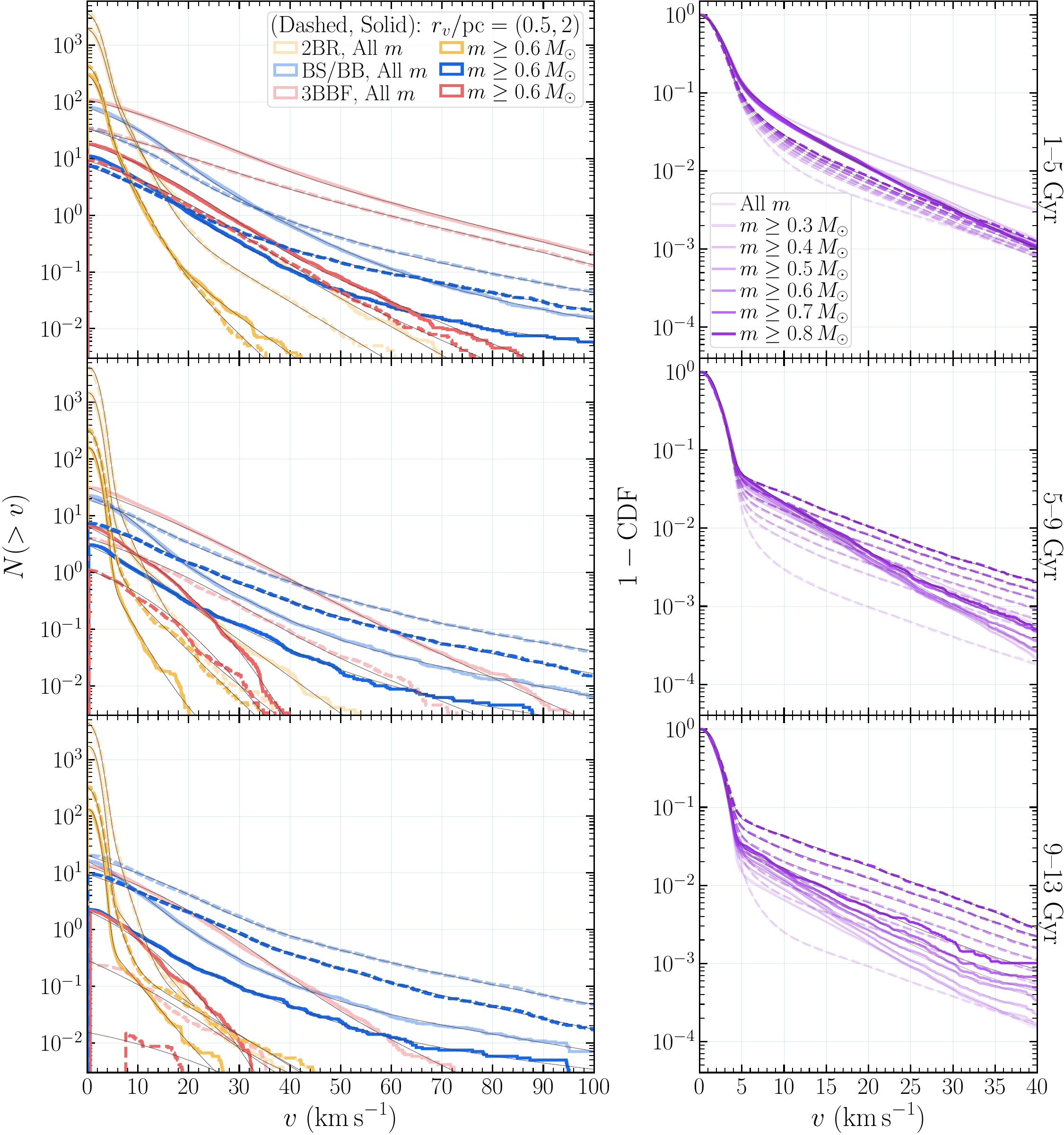}
\caption{Distribution of our stream stars' velocities relative to the GC's orbital velocity, $v=\lvert\vec{v_\star}-\vec{v_c}\rvert$, for several GC age ranges: $1$--$5\,\Gyr$ (top row), $5$--$9\,\Gyr$ (middle), and $9$--$13\,\Gyr$ (bottom). We only count stars with $\lvert\phi\rvert\leq30^\circ$ and combine data from all six simulations for each choice of $r_v$---$0.5\,\pc$ (dashed curves) and $2\,\pc$ (solid). The left panels show the time-averaged number of stream stars with velocities ${>}v$ for several escape mechanisms: two-body relaxation (2BR; yellow), binary--single and binary--binary strong encounters (BS/BB; blue), and three-body binary formation (3BBF; red). The lighter set of curves includes stars regardless of mass, while the darker set counts only stars with $m\geq 0.6\,M_\odot$. The right panels show one minus the overall cumulative density function of $v$, including all escape mechanisms. Line shading again indicates distinct mass cuts, from including all $m$ (lightest) to just stars with $m\geq 0.8\,M_\odot$ (darkest). The prominence of the high-$v$ tail increases with $m$ during and after the GC's core collapse (dashed curves in the lowermost two right panels), precipitated by the loss of most BHs. Thin black curve fits, parameterized in Table~\ref{table:curve_fits_rel}, are plotted atop all curves in the figure, based on stars with $v<100\,\kms$ in the left column (except 2BR; $v<40\,\kms$) and $v<40\,\kms$ in the right column. Equivalent figures for $v$'s component velocities $(v_R,v_\phi,v_Z)$ are in the Appendix.}
\label{fig:vdist_tot}
\end{figure*}

\section{Results} \label{S:results}
We now examine our streams' kinematics, excluding compact objects (assumed unresolved). We also exclude binaries (${\approx}4\%$ of the stream for either $r_v$) to avoid, for now, considering their internal motions, effective observed mass, or selection effects. However, we share some details of the binaries in Section~\ref{S:discussion_binaries} due to interest in how much unresolved binaries may bias observationally inferred radial velocities.

Our streams' velocity distributions are \textit{anisotropic}, so we analyze them via several metrics. This Section focuses on the distribution and dispersion $\sigma_v$ of the absolute relative velocity between each stream star and the GC, $v=\lvert\vec{v_\star}-\vec{v_c}\rvert =\lvert(v_{\star,R},v_{\star,\phi}-v_c,v_{\star,Z})\rvert$, using the Galactocentric cylindrical coordinates defined above. But we also report the distributions of $v$'s individual components $(v_R,v_\phi,v_Z)$, and corresponding dispersions ($\sigma_R,\sigma_\phi,\sigma_Z)$, in the Appendix. Due to the anisotropy, $\sigma_v^2<\sigma_R^2+\sigma_\phi^2+\sigma_Z^2$, and we find $\sigma_R>(\sigma_\phi,\sigma_Z)$ by a constant ${\approx}1.0\,\kms$ for all $m$. Even so, all trends reported in this Section for ($v$, $\sigma_v$) also apply \textit{qualitatively} to $(v_R,v_\phi,v_Z)$ and ($\sigma_R,\sigma_\phi,\sigma_Z)$ in the Appendix. \textit{Hereafter, when we use $\sigma$ with no subscript, we mean the velocity dispersion in streams more generally (any definition, theoretical or observed).}

\subsection{The Velocity Distribution and Escape Mechanisms} \label{S:results_velocity_distribution}
As described earlier, most stars escaping from GCs are unbound by two-body relaxation (2BR) or fluctuating tides. Since our GCs circularly orbit in a smooth Galactic potential, the tides do not fluctuate and 2BR dominates. There are also many types of strong encounters in a GC's core that eject stars more vigorously, contributing a high-$v$ tail to the stream's velocity distribution. The strong ejection mechanisms most relevant to our evolved GCs are binary--single and binary--binary (BS/BB) interactions, and three-body binary formation (3BBF)---a close encounter between three singles that produces a binary, accelerating the leftover single and the new binary's center-of-mass. But the 3BBF rate, $\Gamma\propto n^3m^5\sigma_c^{-9}$, is far more sensitive to star mass $m$ and local velocity dispersion $\sigma_c$ than the local number density $n$. So ejections via 3BBF stall as a GC ages and loses its BHs (massive and slow), despite the core then collapsing to a denser state \citep{Weatherford2023}. The dependence on $m$ and $\sigma_c$ overwhelm the $n$ dependence here, as BHs have $m\sim 20$ times higher than, and $\sigma_c\sim 1/2$, the mean for stars in the core.

Figure~\ref{fig:vdist_tot} details the $v$ distribution in our streams, and the contribution from each escape mechanism. We only count stars with $\lvert \phi\rvert\leq30^\circ$, again combining data from all six simulations for each choice of $r_v$---$0.5\,\pc$ (dashed curves) and $2\,\pc$ (solid). Panel rows correspond to distinct GC age ranges: $1$--$5\,\Gyr$ (top), $5$--$9\,\Gyr$ (middle), and $9$--$13\,\Gyr$ (bottom). The left column shows the number of stream stars, averaged over the age range, that have relative velocities ${>}v$, with color distinguishing escape mechanism---2BR (yellow), BS/BB (blue), and 3BBF (red). The lighter set of curves counts stars of any $m$ while the darker set counts only stars with $m\geq 0.6\,M_\odot$, mimicking magnitude-limited observations.

The left column makes clear that ejecta from 2BR dominate the stream, contributing a dynamically cold core to the velocity distribution with $v\lesssim 5\,\kms$, while rarer ejecta from strong encounters (BS/BB and 3BBF) add a high-$v$ tail to the distribution extending beyond even $v \sim 100\,\kms$. (At any given time, though, only ${\sim}1$ stream member with $\lvert\phi\rvert\leq30^\circ$ has $v\gtrsim 20\,\kms$.) The $v$ distribution of the cold ejecta from 2BR does not depend strongly on star mass $m$ or GC age, in accord with the expectation for a slow diffusive process. The shape of the $v$ distribution from strong BS/BB interactions is also highly consistent across GC age, and BS/BB ejecta are unsurprisingly more numerous in the streams from the denser GCs (dashed curves). Due to gradual ejection of BHs, which occurs faster in denser GCs, there are also far fewer ejecta from 3BBF in the streams at higher GC age and lower $r_v$---even being entirely absent in the stream for $r_v=0.5\,\pc$ by the $9$--$13\,\Gyr$ age range typical of MWGCs.

At early times, though, when many BHs remain in the GC regardless of initial $r_v$, 3BBF actually contributes more to the high-$v$ tail than BS/BB interactions. This appears to be true only when considering ejecta of any mass (light red curves), not when limiting to $m\geq0.6\,M_\odot$ (dark red), but such a conclusion is unreliable. The reason 3BBF overwhelmingly ejects lower-$m$ stars in our models is simply because \texttt{CMC}'s rough prescription for 3BBF \citep{Morscher2013, Morscher2015} automatically pairs the two most-massive bodies in the interaction, following the familiar preference of BS/BB interactions \cite[e.g.,][]{Heggie1975, HeggieHut2003}. But new scattering experiments \citep{Atallah2024} and calculations \citep{GinatPerets2024} show this pairing is actually the \textit{least} likely outcome of 3BBF, so a more realistic 3BBF prescription would not preferentially eject low-$m$ stars. Due to this $m$ bias in our 3BBF ejecta, and since the ejections from 3BBF no longer dominate the high-$v$ tail for GCs of typical age anyway, we henceforth exclude 3BBF ejecta from our analysis.

\floattable
\begin{deluxetable*}{cccccc|ccccc}[ht!]
\tabletypesize{\footnotesize}
\tablewidth{0pt}
\tablecaption{Fitting Parameters Describing the Stream Velocity Distribution\vspace{-0.4cm} \label{table:curve_fits_rel}}
\tablehead{
    \colhead{Velocity} &
    \colhead{Escape} &
    \colhead{$r_v$} &
    \colhead{$t_1$} &
    \colhead{$t_2$} &
    \colhead{$m_L$} &
    \colhead{$\sigma_1$} &
    \colhead{$\sigma_2$} &
    \colhead{$\sigma_3$} &
    \colhead{$a_1$} &
    \colhead{$a_2$} \\[-0.2cm]
    \colhead{Type} &
    \colhead{Mechanism} &
	\colhead{$(\pc)$} &
    \colhead{$(\Gyr)$} &
	\colhead{$(\Gyr)$} &
    \colhead{$(\Msun)$} &
	\colhead{$(\kms)$} &
	\colhead{$(\kms)$} &
	\colhead{$(\kms)$} &
	\colhead{} &
	\colhead{}
}
\startdata
$v$ & All & 0.5 & 1 &  5 & 0.08 &   1.628 $\pm$  0.004 &   2.40 $\pm$   0.01 &  13.69 $\pm$   0.04 &  0.6179 $\pm$  0.0065 &  0.3645 $\pm$  0.0064 \\
$v$ & All & 0.5 & 1 &  5 & 0.3  &   1.603 $\pm$  0.004 &   2.30 $\pm$   0.02 &  10.27 $\pm$   0.02 &  0.6204 $\pm$  0.0066 &  0.3420 $\pm$  0.0065 \\
$v$ & All & 0.5 & 1 &  5 & 0.4  &   1.589 $\pm$  0.003 &   2.38 $\pm$   0.02 &  10.09 $\pm$   0.02 &  0.6244 $\pm$  0.0060 &  0.3345 $\pm$  0.0059 \\
$v$ & All & 0.5 & 1 &  5 & 0.5  &   1.577 $\pm$  0.004 &   2.52 $\pm$   0.02 &  10.17 $\pm$   0.02 &  0.6414 $\pm$  0.0065 &  0.3153 $\pm$  0.0064 \\
$v$ & All & 0.5 & 1 &  5 & 0.6  &   1.586 $\pm$  0.004 &   2.51 $\pm$   0.02 &  10.19 $\pm$   0.02 &  0.6141 $\pm$  0.0063 &  0.3373 $\pm$  0.0062 \\
$v$ & All & 0.5 & 1 &  5 & 0.7  &   1.577 $\pm$  0.005 &   2.63 $\pm$   0.03 &  10.30 $\pm$   0.03 &  0.6251 $\pm$  0.0076 &  0.3239 $\pm$  0.0075 \\
$v$ & All & 0.5 & 1 &  5 & 0.8  &   1.579 $\pm$  0.005 &   2.78 $\pm$   0.04 &  10.57 $\pm$   0.04 &  0.6485 $\pm$  0.0088 &  0.2977 $\pm$  0.0086 \\ \hline
$v$ & All & 0.5 & 5 &  9 & 0.08 &   1.431 $\pm$  0.002 &   1.89 $\pm$   0.03 &  11.49 $\pm$   0.03 &  0.8774 $\pm$  0.0054 &  0.1170 $\pm$  0.0054 \\
$v$ & All & 0.5 & 5 &  9 & 0.3  &   1.419 $\pm$  0.002 &   3.10 $\pm$   0.09 &  12.11 $\pm$   0.07 &  0.9467 $\pm$  0.0023 &  0.0403 $\pm$  0.0022 \\
$v$ & All & 0.5 & 5 &  9 & 0.4  &   1.412 $\pm$  0.002 &   2.76 $\pm$   0.09 &  11.94 $\pm$   0.05 &  0.9299 $\pm$  0.0038 &  0.0513 $\pm$  0.0037 \\
$v$ & All & 0.5 & 5 &  9 & 0.5  &   1.409 $\pm$  0.003 &   4.08 $\pm$   0.15 &  12.75 $\pm$   0.11 &  0.9415 $\pm$  0.0019 &  0.0377 $\pm$  0.0015 \\
$v$ & All & 0.5 & 5 &  9 & 0.6  &   1.407 $\pm$  0.003 &   6.07 $\pm$   0.30 &  14.24 $\pm$   0.43 &  0.9434 $\pm$  0.0012 &  0.0372 $\pm$  0.0011 \\
$v$ & All & 0.5 & 5 &  9 & 0.7  &   1.403 $\pm$  0.003 &   5.42 $\pm$   0.37 &  12.66 $\pm$   0.23 &  0.9301 $\pm$  0.0018 &  0.0346 $\pm$  0.0013 \\
$v$ & All & 0.5 & 5 &  9 & 0.8  &   1.383 $\pm$  0.003 &   9.41 $\pm$   0.18 &  51.15 $\pm$  23.90 &  0.9222 $\pm$  0.0008 &  0.0757 $\pm$  0.0005 \\ \hline
$v$ & All & 0.5 & 9 & 13 & 0.08 &   1.405 $\pm$  0.002 &   1.57 $\pm$   0.01 &  11.58 $\pm$   0.02 &  0.7627 $\pm$  0.0050 &  0.2322 $\pm$  0.0050 \\
$v$ & All & 0.5 & 9 & 13 & 0.3  &   1.398 $\pm$  0.002 &   1.62 $\pm$   0.01 &  11.68 $\pm$   0.02 &  0.7775 $\pm$  0.0057 &  0.2063 $\pm$  0.0057 \\
$v$ & All & 0.5 & 9 & 13 & 0.4  &   1.397 $\pm$  0.002 &   1.67 $\pm$   0.02 &  11.64 $\pm$   0.02 &  0.7826 $\pm$  0.0058 &  0.1939 $\pm$  0.0058 \\
$v$ & All & 0.5 & 9 & 13 & 0.5  &   1.409 $\pm$  0.003 &   1.58 $\pm$   0.02 &  11.64 $\pm$   0.02 &  0.7585 $\pm$  0.0075 &  0.2081 $\pm$  0.0075 \\
$v$ & All & 0.5 & 9 & 13 & 0.6  &   1.401 $\pm$  0.004 &   1.69 $\pm$   0.03 &  11.77 $\pm$   0.03 &  0.7606 $\pm$  0.0100 &  0.1932 $\pm$  0.0100 \\
$v$ & All & 0.5 & 9 & 13 & 0.7  &   1.398 $\pm$  0.005 &   1.63 $\pm$   0.03 &  11.66 $\pm$   0.02 &  0.7381 $\pm$  0.0094 &  0.1939 $\pm$  0.0093 \\
$v$ & All & 0.5 & 9 & 13 & 0.8  &   1.403 $\pm$  0.003 &  11.33 $\pm$   0.24 &  100                &  0.8959 $\pm$  0.0010 &  0.1045 $\pm$  0.0009 \\ \hline \hline
$v$ & All & 2   & 1 &  5 & 0.08 &   1.662 $\pm$  0.004 &   3.11 $\pm$   0.03 &  11.81 $\pm$   0.03 &  0.6089 $\pm$  0.0064 &  0.3016 $\pm$  0.0061 \\
$v$ & All & 2   & 1 &  5 & 0.3  &   1.648 $\pm$  0.004 &   2.61 $\pm$   0.03 &   9.00 $\pm$   0.01 &  0.5985 $\pm$  0.0061 &  0.2927 $\pm$  0.0059 \\
$v$ & All & 2   & 1 &  5 & 0.4  &   1.634 $\pm$  0.004 &   2.67 $\pm$   0.03 &   8.63 $\pm$   0.01 &  0.6015 $\pm$  0.0061 &  0.2843 $\pm$  0.0058 \\
$v$ & All & 2   & 1 &  5 & 0.6  &   1.639 $\pm$  0.004 &   2.87 $\pm$   0.03 &   8.52 $\pm$   0.02 &  0.6025 $\pm$  0.0064 &  0.2877 $\pm$  0.0061 \\
$v$ & All & 2   & 1 &  5 & 0.7  &   1.638 $\pm$  0.006 &   3.09 $\pm$   0.06 &   8.44 $\pm$   0.03 &  0.6132 $\pm$  0.0093 &  0.2752 $\pm$  0.0085 \\
$v$ & All & 2   & 1 &  5 & 0.8  &   1.636 $\pm$  0.006 &   3.42 $\pm$   0.07 &   8.59 $\pm$   0.04 &  0.6372 $\pm$  0.0083 &  0.2559 $\pm$  0.0072 \\ \hline
$v$ & All & 2   & 5 &  9 & 0.08 &   1.396 $\pm$  0.004 &   1.50 $\pm$   0.05 &   9.13 $\pm$   0.01 &  0.8374 $\pm$  0.0094 &  0.1181 $\pm$  0.0094 \\
$v$ & All & 2   & 5 &  9 & 0.3  &   1.373 $\pm$  0.003 &   6.29 $\pm$   0.36 &   8.97 $\pm$   1.59 &  0.9261 $\pm$  0.0012 &  0.0639 $\pm$  0.0122 \\
$v$ & All & 2   & 5 &  9 & 0.4  &   1.367 $\pm$  0.004 &   6.25 $\pm$   0.10 &  15.37 $\pm$   2.60 &  0.9167 $\pm$  0.0011 &  0.0815 $\pm$  0.0007 \\
$v$ & All & 2   & 5 &  9 & 0.5  &   1.378 $\pm$  0.004 &   6.29 $\pm$   0.10 &  16.10 $\pm$   3.58 &  0.9114 $\pm$  0.0013 &  0.0871 $\pm$  0.0008 \\
$v$ & All & 2   & 5 &  9 & 0.6  &   1.369 $\pm$  0.004 &   6.32 $\pm$   0.04 &  100                &  0.9034 $\pm$  0.0010 &  0.0963 $\pm$  0.0010 \\
$v$ & All & 2   & 5 &  9 & 0.7  &   1.385 $\pm$  0.004 &   6.25 $\pm$   0.06 &  25.66 $\pm$   3.67 &  0.8985 $\pm$  0.0012 &  0.0999 $\pm$  0.0009 \\
$v$ & All & 2   & 5 &  9 & 0.8  &   1.394 $\pm$  0.004 &   6.49 $\pm$   0.07 &  31.64 $\pm$   6.20 &  0.9024 $\pm$  0.0011 &  0.0963 $\pm$  0.0009 \\ \hline
$v$ & All & 2   & 9 & 13 & 0.08 &   1.346 $\pm$  0.005 &   0.49 $\pm$   0.56 &   7.82 $\pm$   0.01 &  0.9204 $\pm$  0.0208 &  0.0565 $\pm$  0.0208 \\
$v$ & All & 2   & 9 & 13 & 0.3  &   1.326 $\pm$  0.003 &   6.28 $\pm$   0.04 & 100                 &  0.9578 $\pm$  0.0004 &  0.0420 $\pm$  0.0004 \\
$v$ & All & 2   & 9 & 13 & 0.5  &   1.329 $\pm$  0.004 &   6.09 $\pm$   0.08 &  29.01 $\pm$   3.87 &  0.9440 $\pm$  0.0009 &  0.0549 $\pm$  0.0007 \\
$v$ & All & 2   & 9 & 13 & 0.6  &   1.327 $\pm$  0.005 &   6.81 $\pm$   0.09 & 100                 &  0.9430 $\pm$  0.0009 &  0.0565 $\pm$  0.0008 \\
$v$ & All & 2   & 9 & 13 & 0.7  &   1.326 $\pm$  0.005 &   6.94 $\pm$   0.10 & 100                 &  0.9377 $\pm$  0.0011 &  0.0617 $\pm$  0.0009 \\
$v$ & All & 2   & 9 & 13 & 0.8  &   1.346 $\pm$  0.006 &   8.20 $\pm$   0.23 & 100                 &  0.9450 $\pm$  0.0013 &  0.0544 $\pm$  0.0009 \\
\enddata
\vspace{-0.1cm}
\tablecomments{\footnotesize This table will be accessible in its entirety and in machine-readable format in the online version of this publication, including the fitting parameters for both the $v$ distribution (Figure~\ref{fig:vdist_tot}'s right column, shown here) \textit{and} the individual contributions from each escape mechanism (Figure~\ref{fig:vdist_tot}'s left column), plus the same information for the distributions of $v$'s components $(v_R,v_\phi,v_Z)$. The first two columns indicate the velocity type (``v" for $v$, ``r" for $v_R$, ``phi" for $v_\phi$, and ``z" for $v_Z$) and the escape mechanisms included in the fit (2BR, BS/BB, 3BBF, or ``All" for the combination of all three). The 3rd--6th columns list the $r_v$ of contributing GC simulations, the age range $(t_1,t_2)$ of contributing stream snapshots, and the minimum mass ($m_L$) stars must have to be counted. The remaining columns are the fitted parameters from Equations~(\ref{eq:mee_fit}) and (\ref{eq:ne_fit})---or Equations~(\ref{eq:nee_fit}) and (\ref{eq:ee_fit}) when fitting the distributions of $(v_R,v_\phi,v_Z)$---each accompanied by an uncertainty (the standard deviation from each fit's covariance matrix). In accord with these equations, the columns for $\sigma_3$, $a_2$, and their corresponding uncertainties have \revision{NULL} entries in the machine-readable table online when fitting the contributions to the velocity distribution from BS/BB or 3BBF ejecta. Maxed-out values of 100 in the columns for $\sigma_3$ and its uncertainty reflect our fitting constraint that $(\sigma_1,\sigma_2,\sigma_3)$ must be less than $100\,\kms$. Similarly, values of 100 in the $a_2$ column and its uncertainty mean $\sigma_2$ and $\sigma_3$ were degenerate. Either case indicates a one-component exponential tail is satisfactory on its own.}
\end{deluxetable*} 

Figure~\ref{fig:vdist_tot}'s right column shows the survival function (one minus the cumulative density function) of $v$ including all escape mechanisms. Line shading indicates mass cut, from including all $m$ (lightest shade) to just stars with $m\geq 0.8\,M_\odot$ (darkest). These panels reveal the high-$v$ tail's prominence grows once the core begins to contract ($t\approx5\,\Gyr$ in the GCs born with $r_v=0.5\,\pc$; Figure~\ref{fig:macro_evolution}). By this time the tail's prominence also rises with increasing $m$, due to increasing stellar mass segregation as GCs age and eject their BHs (Figure~\ref{fig:mass_segregation}). Note that significant $m$ dependence in the $v$ distribution has not yet set in even by $9$--$13\,\Gyr$ for the less-dense GCs born with $r_v=2\,\pc$, as the heating from BHs remaining in the core resists stellar mass segregation \citep{Mackey2008, Alessandrini2016, Peuten2016, Webb2017, Weatherford2018, Weatherford2020}. So, while BH-catalyzed 3BBF promotes high-$v$ escape early in the GC's life (upper panel), by late times it is actually the \textit{loss} of BHs---and the ensuing density contraction and boost to the BS/BB rate---that promotes high-$v$ escape! This runs somewhat counter to the intuition that BHs in GCs lead to stronger ejections---e.g., in the extreme case of a central intermediate-mass BH \citep{Pfahl2005, BaumgardtMakino2005, Gualandris2007}.

\subsection{Fitting the Velocity Distributions} \label{S:results_curve_fitting}
To quantify the $v$ distribution and aid comparisons to our data, we apply a least-squares fit to all curves in Figure~\ref{fig:vdist_tot}, for robustness based only on stars with $v\leq40\,\kms$ (or $v\leq100\,\kms$ for ejecta from BS/BB or 3BBF in the left column). These fits appear as thin black lines atop every curve. We find 2BR's contribution to $v$ is well fit by a Maxwellian distribution plus a one- or two-component exponential tail\footnote{Our exponential tail for 2BR does not necessarily reflect the true dynamics of relaxation since our classification of escape mechanisms in \texttt{CMC} cannot distinguish whether stars were kicked to sufficient energy for escape by \texttt{CMC}'s actual 2BR algorithm or (much more rarely) by purely \textit{ad hoc} dumping of excess energy to neighboring stars to conserve global energy \citep{Stodolkiewicz1982,CMCRelease}. Furthermore, the \cite{Chandrasekhar1942} theory of relaxation underpinning \texttt{CMC}'s 2BR algorithm applies only to the cumulative effect of many weak encounters, so does not consider strong two-body encounters, which might otherwise be expected to contribute a small exponential tail. \texttt{CMC}'s 2BR algorithm may lead to a qualitatively similar tail by accident, but not by design.}; i.e., the probability density function for $v$ has the form
\begin{equation} \label{eq:mee_fit}
\begin{aligned}
P_1(v|&\sigma_1,\sigma_2,\sigma_3,a_1,a_2) =\\
&a_1\sqrt{\frac{2}{\pi}}\frac{v^2}{\sigma_1^3} e^{-\frac{v^2}{2\sigma_1^2}} + \frac{a_2}{\sigma_2}e^{-\frac{v}{\sigma_2}} + \frac{(1-a_1-a_2)}{\sigma_3}e^{-\frac{v}{\sigma_3}},
\end{aligned}
\end{equation}
where the fitted parameters are $(\sigma_1,\sigma_2,\sigma_3,a_1,a_2)$. The contribution to $v$ from BS/BB strong encounters, and the overall $v$ distribution (including all mechanisms; Figure~\ref{fig:vdist_tot}'s right column) also take the form of Equation~(\ref{eq:mee_fit}), but with different fitting parameter values. In all of the above cases, the Maxwellian term becomes Gaussian when fitting $v$'s individual component velocities $(v_R,v_\phi,v_Z)$ in the Appendix, in accord with kinetic theory.

However, the contribution to $v$ from 3BBF is better fit by a Gaussian with an exponential tail (or a two-component exponential distribution for $v$'s individual components in the Appendix). This has a probability density function of the form,
\begin{equation} \label{eq:ne_fit}
P_2(v|\sigma_1,\sigma_2,a_1) = \frac{a_1}{\sigma_1} \sqrt{\frac{2}{\pi}} e^{-\frac{v^2}{2\sigma_1^2}} + \frac{1-a_1}{\sigma_2} e^{-\frac{v}{\sigma_2}}.
\end{equation}

To increase accuracy in the tail, we perform all fits on the logarithm of the survival function (one minus the cumulative density function) of $v$, sampled every $0.1\,\kms$.

\subsection{Velocity Dispersion} \label{S:results_velocity_dispersion}
\begin{figure*}[ht!]
\centering
\includegraphics[width=\linewidth]{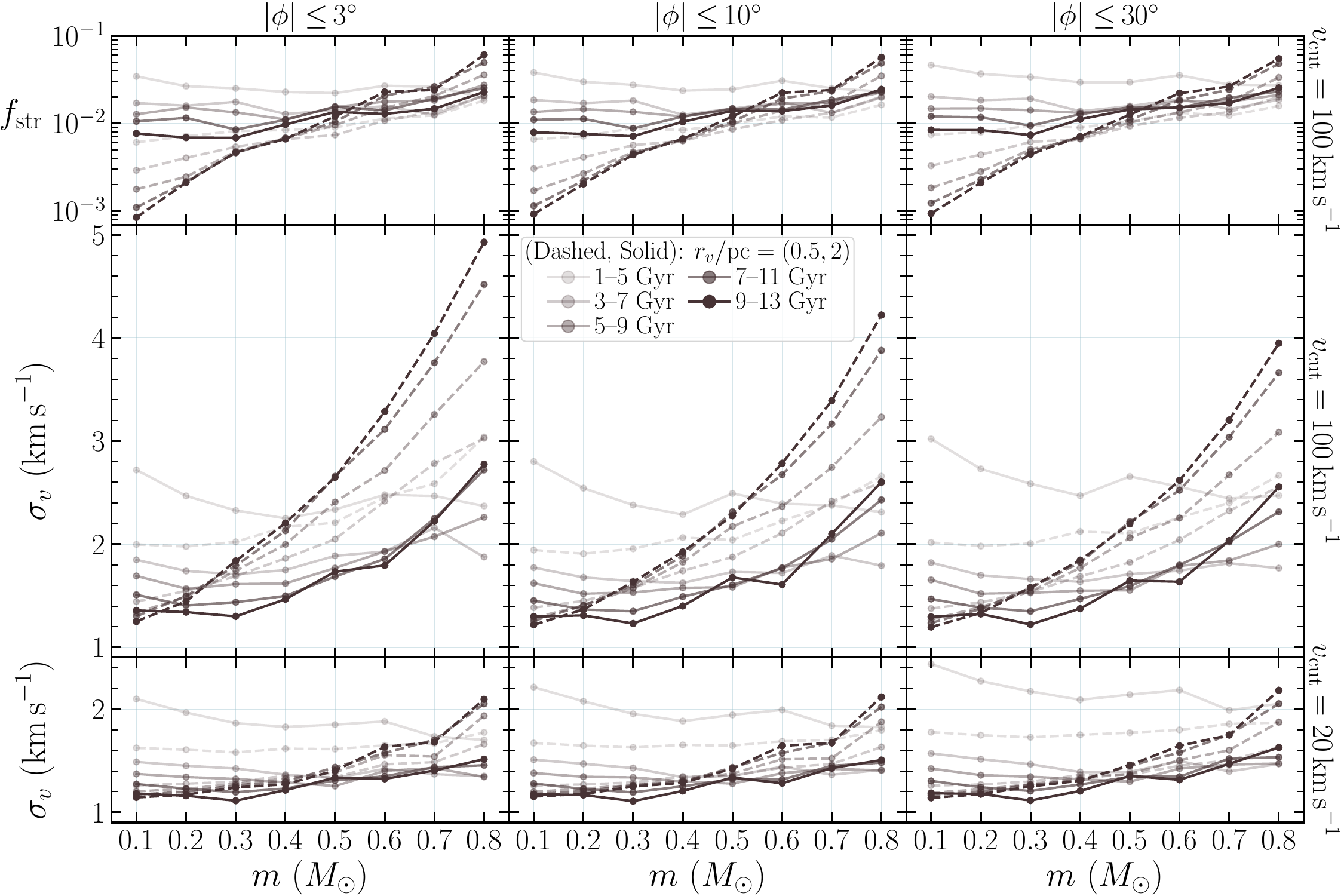}
\caption{The fraction $f_{\rm str}$ of stream members ejected by BS/BB encounters (top row) and three-dimensional velocity dispersion $\sigma_v$ (lower rows) in our stellar streams versus star mass $m$. As in Figure~\ref{fig:mass_segregation}, we use non-overlapping mass bins of width $0.1\,\Msun$ and combine the data from all six independent simulations for each choice of $r_v$---$0.5\,\pc$ (dashed curves) and $2\,\pc$ (solid curves). Shading again indicates time averages across several different $4\,\Gyr$-wide intervals in GC age (see legend). The upper two rows of panels include stars for which each $v$ component $(v_R,v_\phi,v_Z)\leq100\,\kms$---nearly all $v$ but filtering out several extreme outliers. The bottom row includes only stars with $(v_R,v_\phi,v_Z)\leq20\,\kms$, similar to kinematic cuts imposed in observations to help identify stream members. Each column of panels imposes a different angular cut on stream membership: only including stars with $\lvert\phi\rvert\leq3^\circ$ (left), $\lvert\phi\rvert\leq10^\circ$ (center), and $\lvert\phi\rvert\leq30^\circ$ (right, most typical of observed stream lengths). The $m$--$f_{\rm str}$ correlation in the top row causes an $m$--$\sigma_v$ correlation in the lower panels, albeit reduced under the more stringent velocity cut. Equivalent figures for $v$'s component velocities $(v_R,v_\phi,v_Z)$ are in the Appendix.}
\label{fig:vdisp_all_snaps_tot}
\end{figure*}

The increasing prominence of the $v$ distribution's tail at higher $m$ naturally leads to higher $\sigma_v$ (recall this is the true overall dispersion in $v$, versus fitting parameters $\sigma_1,\sigma_2,\sigma_3$). This is evident in Figure~\ref{fig:vdisp_all_snaps_tot}'s lower rows, which plot $\sigma_v$ versus $m$, while the top row shows the fraction $f_{\rm str}$ of stars present in the stream that were ejected by strong BS/BB encounters. As in Figure~\ref{fig:mass_segregation}, $f_{\rm str}$ and $\sigma_v$ in each $m$ bin are averaged across several age ranges (distinguished by line shading) and combine data from all six simulations for each choice of $r_v$---$0.5\,\pc$ (dashed curves) and $2\,\pc$ (solid). To show how $\sigma_v$ depends on position along the stream, each column imposes a different angular cut on membership: counting only stars with $\lvert\phi\rvert\leq3^\circ$ (left), $\lvert\phi\rvert\leq10^\circ$ (center), and $\lvert\phi\rvert\leq30^\circ$ (right).

The upper two rows of Figure~\ref{fig:vdisp_all_snaps_tot} count stars for which each $v$ component $(v_R,v_\phi,v_Z)\leq100\,\kms$, essentially all $v$ but filtering out a few extreme outliers. These rows show $f_{\rm str}$ and $\sigma_v$ correlate increasingly steeply with $m$ as the stream's progenitor GC ages (darker curves). The trend is steepest once the loss of most BHs in the denser GC allows its core to collapse, promoting mass segregation (Figure~\ref{fig:mass_segregation}), which explains why the highest-$m$ stars, deepest in the GC, are the most likely to be ejected by strong encounters. The enhanced mass segregation and encounter rates after core collapse in the denser GC cause $f_{\rm str}$ to increase exponentially from ${\sim}10^{-3}$--$10^{-1}$ over the mass range $m=0.1$--$0.8\,\Msun$.

Unlike for $m$--$f_{\rm str}$, the $m$--$\sigma_v$ relation mildly depends on the angular cut on stream membership. For typical GC ages ($9$--$13\,\Gyr$), $\sigma_v$ in the core-collapsed GCs ($r_v=0.5\,\pc$) increases sharply over $m=0.1$--$0.8\,\Msun$, from $1.2$--$4.0\,\kms$ when counting all stars with $\lvert\phi\rvert\leq30^\circ$ (right column), and from $1.2$--$4.9\,\kms$ when only counting stars with $\lvert\phi\rvert\leq3^\circ$ (left). The increase is similar, albeit smaller, for the non-core-collapsed GCs ($r_v=2\,\pc$), respectively reaching up to $2.6\,\kms$ and $2.8\,\kms$. This weak $\lvert\phi\rvert$--$\sigma_v$ anticorrelation is due to more-energetic ejecta passing through wider openings in the tidal equipotential surfaces enclosing the GC. Such ejecta are less collimated into the GC's tidal tails, and their escape vectors will not necessarily align with the stream for much distance \citep[Figure~11 of][]{Weatherford2024}. More-energetic stars thus exit our stream selection window earlier, reducing $\sigma_v$ at higher $\lvert\phi\rvert$. The reduction is less notable at low $m$ since these stars are rarely ejected by strong encounters.

\revision{We also checked that our results are not overly sensitive to the \textit{width} of the stream selection window along either the $X$ or $Z$ axis. When including only stars with $(v_R,v_\phi,v_Z)< 20\,\kms$, the values of $\sigma$ for each $m$ bin in Figure~\ref{fig:vdisp_all_snaps_tot} increase/decrease by ${\lesssim}0.4\,\kms$ when halving/doubling the width of either selection criterion ($\lvert X\rvert\leq X_{\rm \lim}=0.05R_g$ and $\lvert Z\rvert\leq Z_{\rm \lim}=0.02R_g$). Halving $X_{\rm lim}$ cuts out most of the stream (Figure~\ref{fig:density_map}) while doubling it results in a more permissive selection than typically applied observationally, even when studying the GD-1 cocoon \citep[e.g.,][]{Valluri2025}. In any case, these shifts are most significant for the highest-$m$ bins since those stars dominate the stream periphery due to their higher $\sigma$---a form of mass segregation perpendicular to the stream centerline.}

The $m$--$\sigma$ correlation in our streams, especially from core-collapsed GCs, is steep enough to be a significant selection effect when measuring $\sigma$ observationally. Yet observers often rely on kinematics to help identify stream members, commonly excluding stars whose velocities differ from the stream's local mean by more than a few times the GC's internal $\sigma_c$. Our inclusion of stars almost irrespective of $v$ is thus an idealized scenario. To mimic kinematic membership cuts often applied to stream observations, the bottom row of Figure~\ref{fig:vdisp_all_snaps_tot} shows the $m$--$\sigma_v$ correlation when limiting to stars with $(v_R,v_\phi,v_Z)\leq20\,\kms$. This greatly dampens the $m$--$\sigma_v$ correlation, with $\sigma_v$ now only increasing from $1.2$--$2.2\,\kms$ for $r_v=0.5\,\pc$, or from $1.2$--$1.6\,\kms$ for $r_v=2\,\pc$. Even so, this remains a significant increase given the low $\sigma\lesssim3\,\kms$ in many observed streams from GCs (see Section~\ref{S:discussion_observed_dispersions}).

\begin{figure*}[ht!]
\centering
\includegraphics[width=0.985\linewidth]{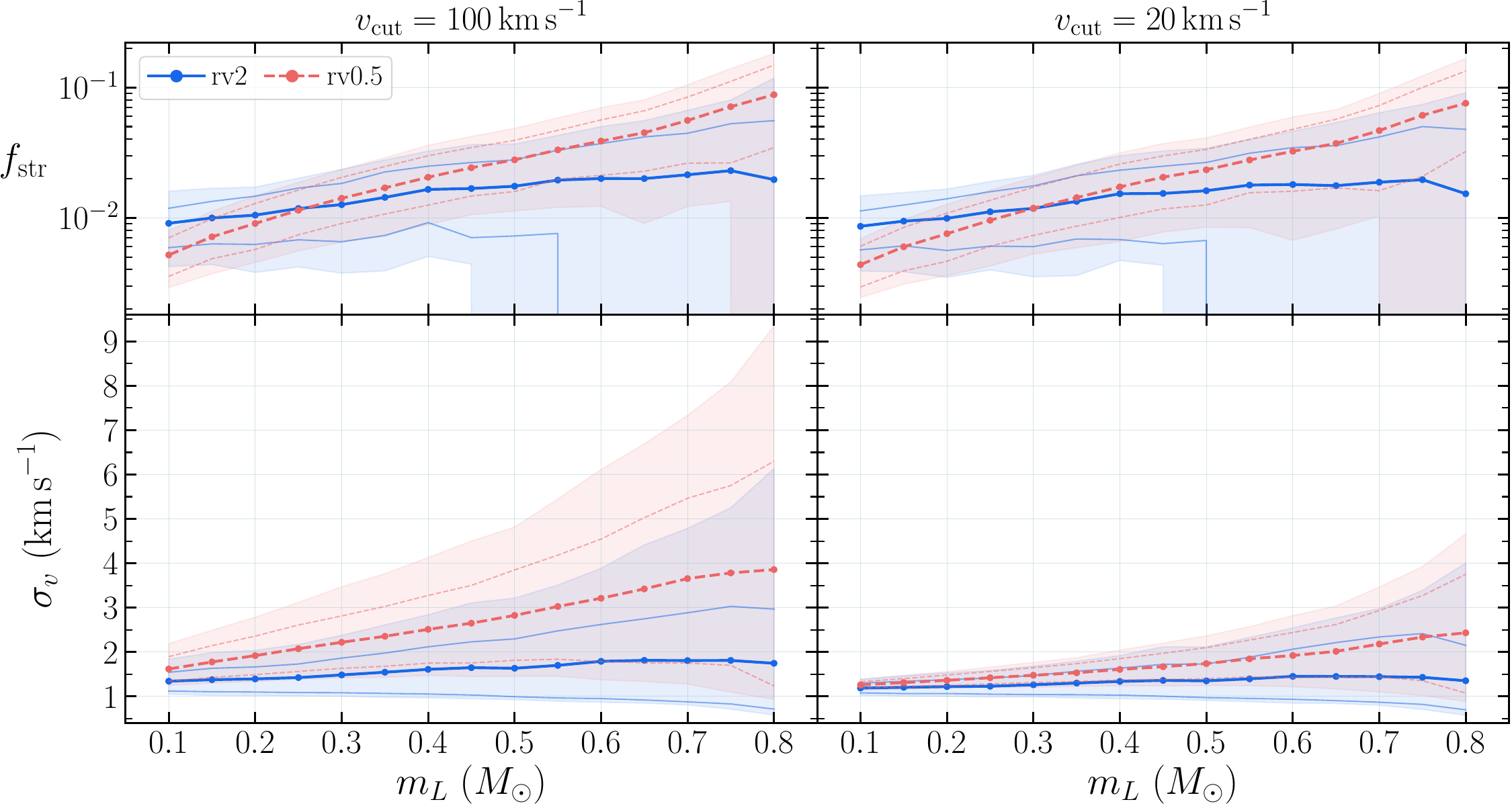}
\caption{The fraction $f_{\rm str}$ of stars ejected by strong BS/BB encounters (top row) and velocity dispersion $\sigma_v$ (lower row) in our streams, this time counting all stars with mass $m\geq m_L$. Unlike in Figure~\ref{fig:vdisp_all_snaps_tot}, we measure $f_{\rm str}$ and $\sigma_v$ \textit{individually} for each stream snapshot with ages $11$--$13\,\Gyr$, and plot the mean of these measurements across the snapshots from all six simulations for each choice of $r_v$---$0.5\,\pc$ (red dashed curve) and $2\,\pc$ (blue solid curve). For each $r_v$, the random fluctuations in $f_{\rm str}$ and $\sigma_v$ over the snapshots (over time) are characterized by the pair of thinner curves (16th--84th percentile range) and shaded regions (5th--95th percentile range) with matching colors. The left column of panels counts all stream members with $(v_R,v_\phi,v_Z)\leq 100\,\kms$, while the right column counts only those with $(v_R,v_\phi,v_Z)\leq 20\,\kms$. Equivalent figures for $v$'s component velocities $(v_R,v_\phi,v_Z)$ are in the Appendix.}
\label{fig:vdisp_indiv_snaps_tot}
\end{figure*}

Another practical consideration affecting the $m$--$\sigma$ correlation's observability in real GC streams are statistical fluctuations in $\sigma$ over time due to few ejecta from strong encounters. To limit noise in our plots so far, we have stacked data from multiple GC simulations, and hundreds of snapshots in time each, before computing $\sigma$. While the snapshots are not statistically independent due to many duplicate appearances of each star,
this approach is analogous to collating data on ${\approx}30$ streams from near-identical GCs, which is observationally infeasible. An approach more comparable to stream observations is to instead compute $\sigma$ \textit{individually} for each snapshot, selecting all stars more massive than some threshold $m_L$ representing the limit of detectability in a given cluster with a given telescope (${\approx}0.5\,\Msun$ with \textit{Gaia} for even the nearest core-collapsed GC, NGC~6397). We do so in Figure~\ref{fig:vdisp_indiv_snaps_tot} for $f_{\rm str}$ (top row) and $\sigma_v$ (bottom row), plotting the mean of the snapshot-specific measurements across all snapshots with ages $11$--$13\,\Gyr$ from all six simulations for each choice of $r_v$---$0.5\,\pc$ (red dashed curve) and $2\,\pc$ (blue solid curve). We characterize the fluctuations in these quantities over time via the accompanying pairs of thinner curves (16th--84th percentile range) and shaded regions (5th--95th percentile range) of matching color. The left column counts stars with $(v_R,v_\phi,v_Z)\leq 100\,\kms$, while the right column again mimics observational selection by limiting to stars with $(v_R,v_\phi,v_Z)\leq 20\,\kms$.

The correlation between $m_L$ and mean $\sigma_v$ in Figure~\ref{fig:vdisp_indiv_snaps_tot} is very similar to that between $m$ bin and $\sigma_v$ in Figure~\ref{fig:vdisp_all_snaps_tot}. For $r_v=0.5\,\pc$ and over the range $m_L=0.1$--$0.8\,\Msun$, mean $\sigma_v$ increases almost linearly from $1.6$--$3.9\,\kms$, or $1.3$--$2.5\,\kms$ when limiting to stars with $(v_R,v_\phi,v_Z)\leq 20\,\kms$. But the scatter in $\sigma_v$ over time is high, especially at higher $m_L$, due to the shrinking sample size. While the streams at any given time contain several hundred stars with $m\geq0.6\,\Msun$  (Figure~\ref{fig:vdist_tot}), only a few ($r_v=2\,\pc$) to a few tens ($r_v=0.5\,\pc$) of these are ejecta from strong encounters, which have an outsize influence on $\sigma_v$. For example, the 5th--95th percentile range of $\sigma_v(m\geq0.6\,\Msun)$ spans $1.4$--$6.1\,\kms$ for $r_v=0.5\,\pc$ ($0.9$--$3.9\,\kms$ for $r_v=2\,\pc$) depending on when the stream is ``observed" and which specific ejecta are in the selection window at that time. These fluctuations are also smaller---$1.2$--$2.8\,\kms$ ($0.9$--$2.6\,\kms$)---when limiting to stars with $(v_R,v_\phi,v_Z)\leq 20\,\kms$. A slight drop in $f_{\rm str}$ and $\sigma_v$ at $m_L\approx 0.8\,\Msun$  for $r_v=2\,\pc$ is due to the typical sample size above this mass being so small (${\approx}$30) that the sample in most snapshots contains no strong ejecta at all.

The large scatter in $\sigma$ over time for high $m_L$ has several important implications. (1) A statistically robust observational verification of the $m$--$\sigma$ correlation in streams from MWGCs requires stacking measurements from multiple streams. (2) It is plausible that observations of streams from GCs limited to stars near and above the main-sequence turnoff will sometimes measure $\sigma\gtrsim3\,\kms$, even in the absence of any heating from DM subhalos or other MW substructure, any pre-processing in a parent halo, or any spurious enhancement from, e.g., unresolved binaries. (3) Observations of $\sigma$ in any \textit{individual} stream, typically magnitude-limited to $m\gtrsim 0.6\,\Msun$ or higher, are relatively low-precision measures of the heating applied to the stream by MW substructure, or of parent halo properties, though $\sigma$ can still rule out causes of especially large $\sigma$, e.g., origins in massive cuspy parent halos. Precise predictions for the properties of DM or parent halos would thus require simultaneous consideration of $\sigma$ observed in many different streams. This situation is not new, as the amount of heating applied to a stream by substructure also varies considerably depending on the stream's orbit and encounter history \citep[e.g.,][]{Carlberg2024b}.

\begin{figure*}[ht!]
\centering
\includegraphics[width=0.99\linewidth]{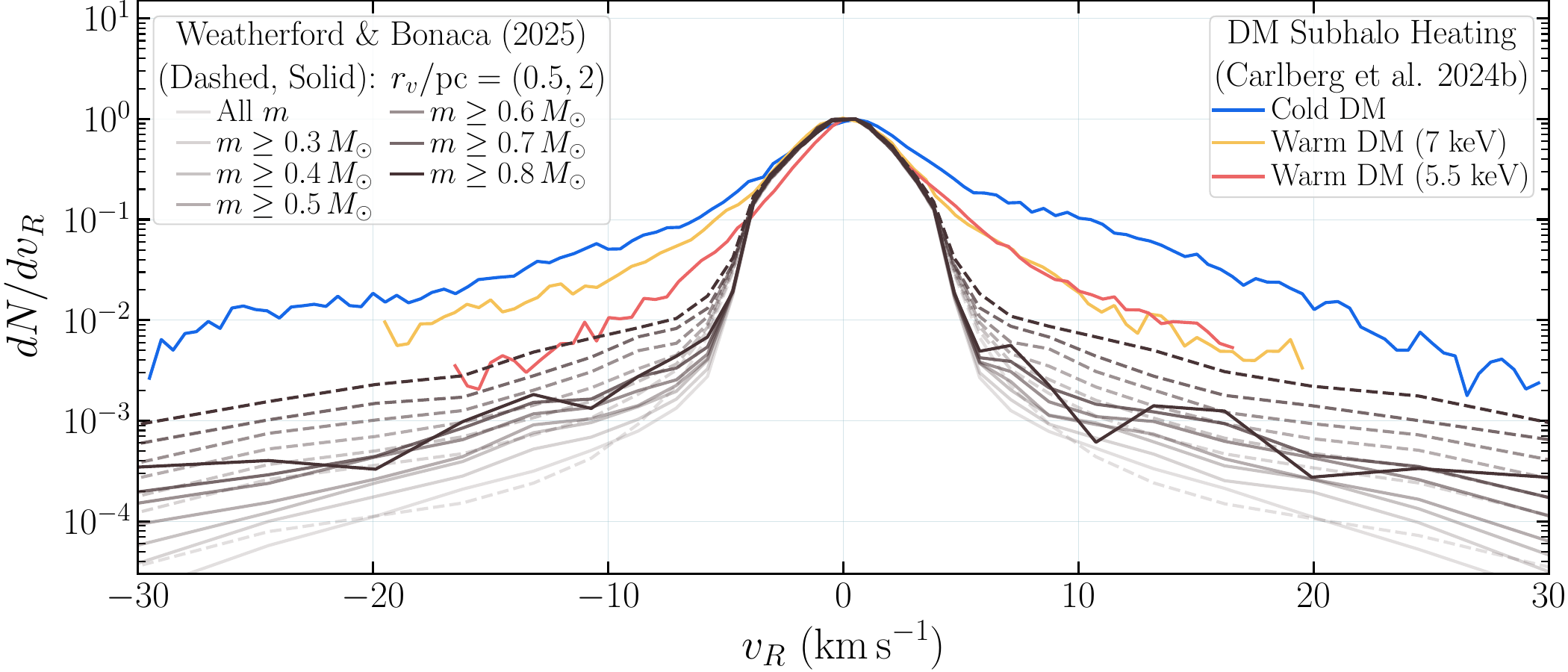}
\caption{Probability density function, normalized to a peak of one, for the star velocities in our streams---this time in terms of Galactocentric radial velocity $v_R$. We only count stars with $\lvert \phi\rvert\leq30^\circ$ and stack stream snapshots in the age range $11$--$13\,\Gyr$ from all six simulations for each choice of $r_v$---$0.5\,\pc$ (dashed black curves) and $2\,\pc$ (solid black). Black line shading again indicates different mass limits, from including all $m$ (lightest) to just stars with $m\geq 0.8\,M_\odot$ (darkest). Additionally, we plot the $v_R$ distributions for GC streams after heating by DM subhalos, predicted by \cite{Carlberg2024b}---their Figure~10. Each of these distributions corresponds to heating by subhalos in Galaxy simulations assuming several different DM models: canonical cold DM (blue curve) and warm DM with particle masses of $7\,{\rm keV}$ (yellow) or $5.5\,{\rm keV}$ (red).}
\label{fig:vdist_Carlberg}
\end{figure*}

\vspace{1cm}
\section{Discussion} \label{S:discussion}
We now discuss the impact of the $m$--$\sigma$ correlation from strong encounters on stream kinematics versus other factors affecting stream $\sigma$ (Section~\ref{S:discussion_other_factors}), compare to $\sigma$ in observed streams (Section~\ref{S:discussion_observed_dispersions}), describe the anticipated impact of fluctuating tides with a cursory comparison to direct $N$-body models (Section~\ref{S:discussion_nbody}), and elaborate on properties of binaries in our streams (Section~\ref{S:discussion_binaries}).

\subsection{Comparison to Other Velocity Dispersion Enhancers} \label{S:discussion_other_factors}
As described in Section~\ref{S:intro}, recent studies have also modeled the impacts on stream kinematics from heating by DM subhalos, which broaden the stream velocity distribution. This heating yields a dominant cold component (Gaussian $\sigma \sim 1$--$2\,\kms$) with broad exponential wings whose dispersion is sensitive to the DM model \citep[$6\,\kms$ for cold DM and $3$--$4\,\kms$ for 5.5--7~keV warm DM;][]{Carlberg2024b}. These wings are qualitatively similar to the exponential tails contributed by strong encounters (not considered in the above study), motivating a direct comparison.

In Figure~\ref{fig:vdist_Carlberg}, we compare our stream velocity distributions (black curves) to those predicted by \cite{Carlberg2024b}---their Figure~10---in substructured cosmological simulations under several DM assumptions (colored curves). To compare on even footing, we now show the probability distribution (normalized to a peak of one) for the Galactocentric radial velocity $v_R$ of stream stars. We find that while the exponential tails due to strong encounters are typically ${\approx}1.5$ orders of magnitude lower than those expected under cold DM (CDM) cosmology, they have comparable prominence to the tails under warm DM cosmology when limiting to stars $m\gtrsim 0.8\,\Msun$. So while the assumed Galactic potential (especially substructure) in each study is quite different, Figure~\ref{fig:vdist_Carlberg} suggests that strong encounters and the $m$--$\sigma$ correlation may be relevant to DM inference, especially for streams from core-collapsed GCs. Observers should at least expect \textit{some} high-$v$ outliers in streams due to strong ejections alone.

\revision{The enhancement of stream $\sigma$ by strong encounter ejecta may also rival heating by other alternatives to CDM, such as ultralight DM \citep[ULDM; e.g.,][]{Hu2000,Hui2017} or some forms of self-interacting DM \citep[SIDM; e.g.,][]{SpergelSteinhardt2000}. ULDM would consist of non-interacting bosons with masses $m_b$ low enough (${<}10^{-20}\,\eV$) to have kiloparsec-scale de Broglie wavelengths, suppressing small-scale DM substructure. Some velocity-dependent SIDM models also suppress small-scale DM substructure to a lesser degree \citep{Nadler2025}. In both cases, one may expect the reduced subhalo population to heat streams less than CDM. However, low-mass SIDM subhalos tend to undergo core collapse, making them denser than in CDM. Since more compact subhalos more strongly perturb stellar streams \citep[e.g.,][]{Nibauer2025b}, it is then unclear---or at least model-dependent---whether streams in SIDM halos should be hotter or colder than in CDM halos. In the case of ULDM, meanwhile, wave-like fluctuations in the halo density profile also heat streams, at a rate scaling inversely with $m_b$: $d\sigma/dt \sim (1\,\kms) (m_b/10^{-22}\,\eV)^{-1.5}\,\Gyr^{-1}$ \citep{AmoriscoLoeb2018, Dalal2021}. This would outpace heating from CDM (and overpredict observed $\sigma$) for $m_b \lesssim 10^{-22}\,{\rm eV}$. But since observations including the Ly $\alpha$ forest \citep{RogersPeiris2021}, ultrafaint dwarf properties \citep{DalalKravtsov2022}, and strong gravitational lensing \citep{Laroche2022} rule out $m_b \lesssim 10^{-20}$, \textit{viable} ULDM models would heat streams less than CDM. So the enhancement of stream $\sigma$ by strong encounter ejecta may rival heating from both warm DM and ULDM, and perhaps even some SIDM models.}

Also keep in mind that many other effects enhance stream $\sigma$, and heating from DM subhalos (or an ex-situ origin) is one of the strongest. The $m$--$\sigma$ correlation from strong encounters may inflate inferred $\sigma$ as much as some of these other effects, even if subhalo heating dominates overall. The other effects inflating $\sigma$ in streams include heating by baryonic MW substructures---e.g., giant molecular clouds \citep{Amorisco2016}, the MW's bar, spiral arms, and disk \citep{Hattori2016, Pearson2019, Banik2019, CarlbergAgler2023}, infalling MW satellites \citep{Garavito-Camargo2019}, and other GCs \citep{DokeHattori2022, Ferrone2025}. Stream $\sigma$ based on radial velocities can also be spuriously inflated by the internal motions of unresolved binaries \citep{Li2019b,Conroy2019} and gravitational redshift's dependence on star mass and radius \citep{Loeb2014}, though the expected enhancement from each is ${\lesssim}0.5\,\kms$.

\setlength{\extrarowheight}{0.09cm}
\floattable
\begin{deluxetable*}{
    >{\raggedright\arraybackslash}m{0.074\textwidth}
    >{\raggedright\arraybackslash}m{0.067\textwidth} |
    >{\raggedright\arraybackslash}m{0.135\textwidth}
    >{\centering  \arraybackslash}m{0.029\textwidth}
    >{\centering  \arraybackslash}m{0.030\textwidth}
    >{\centering  \arraybackslash}m{0.084\textwidth}
    >{\centering  \arraybackslash}m{0.051\textwidth} |
    >{\raggedright\arraybackslash}m{0.376\textwidth}}
\tabletypesize{\footnotesize}
\tablewidth{\textwidth}
\tablecolumns{8}
\tablecaption{Observed Velocity Dispersions of 14 Streams From Globular Clusters \label{table:observed_vdisps}}
\tablehead{
    \colhead{Stream} &
    \colhead{Progenitor} &
    \colhead{Reference} &
    \colhead{$N$} &
    \colhead{$\sigma$} &
    \colhead{$\sigma$} &
    \colhead{$v_{\rm cut}$} &
    \colhead{Comments}\\[-0.2cm]
    \colhead{Name} &
    \colhead{Cluster} &
    \colhead{} &
    \colhead{} &
    \colhead{Type} &
    \colhead{$(\kms)$} &
    \colhead{$(\kms)$} &
    \colhead{}
}
\startdata 
Pal~5      & Pal~5    & \parbox[m]{\linewidth}{\vspace{0.08cm}\raggedright\cite{Odenkirchen2009}\vspace{0.08cm}} & \vspace{0.06cm}\makecell{15\\17} & RV & \makecell{2.2\\4.7} & \makecell{${\approx}12$\\${\approx}40$} & \parbox[m]{\linewidth}{\vspace{0.08cm}Includes only giant stars.\vspace{0.08cm}} \\[0.07cm]
Pal~5      & Pal~5    & \cite{Kuzma2015}       & 47 & RV & $2.1 \pm 0.4$ & 17.5 & \parbox[m]{\linewidth}{\vspace{0.08cm}Includes only giant stars.\vspace{0.08cm}} \\[0.07cm] \hline
M2         & M2       & \cite{Grillmair2022}   & \vspace{0.08cm}\makecell{20\\60} & PM & \makecell{$3.8\pm0.6$\\$8.0\pm1.0$} & NA & \parbox[m]{\linewidth}{\vspace{0.08cm}For stars ${\lesssim}1$ \textit{Gaia} $G$-band magnitude below MS turnoff.\vspace{0.08cm}} \\[0.07cm] \hline
NGC~0288   & NGC~0288 & \cite{Grillmair2025}   & \vspace{0.07cm}\makecell{44\\[0.12cm]57\\[0.07cm]} & PM & \makecell{$2.47^{+0.75}_{-0.39}$\\[0.07cm]$3.04^{+0.56}_{-0.34}$\\[0.07cm]} & NA  & \parbox[m]{\linewidth}{\vspace{0.08cm}For stars ${\lesssim}1$ \textit{Gaia} $G$-band magnitude below MS turnoff. Leading and trailing tails, respectively.\vspace{0.08cm}}\\ \hline
Fj\"{o}rm  & M68      & \cite{Malhan2022b}     & --  & PM & $1.86^{+1.66}_{-1.29}$ & NA & \parbox[m]{\linewidth}{\vspace{0.08cm}For stars ${<}2.6$ \textit{Gaia} $G$-band magnitudes below MS turnoff.\vspace{0.08cm}} \\[0.07cm] \hline
Gj\"{o}ll  & NGC~3201 & \cite{Malhan2022b}     & --  & PM & $2.19^{+2.25}_{-1.53}$ & NA & \parbox[m]{\linewidth}{\vspace{0.08cm}For stars ${<}3.4$ \textit{Gaia} $G$-band magnitudes below MS turnoff.\vspace{0.08cm}} \\[0.07cm] \hline
Sylgr      & NGC~5024 & \cite{Malhan2022b}     & --  & PM & $1.75^{+1.83}_{-1.23}$ & NA & \parbox[m]{\linewidth}{\vspace{0.08cm}For stars ${<}2.4$ \textit{Gaia} $G$-band magnitudes below MS turnoff.\vspace{0.08cm}} \\[0.07cm] \hline
Phlegethon & --       & \cite{Malhan2022b}     & --  & PM & $1.12^{+1.23}_{-0.78}$ & NA &  \\[0.07cm] \hline
GD-1       & --       & \cite{Koposov2010} & 27  & RV & ${<}3$ & NA & \parbox[m]{\linewidth}{\vspace{0.08cm}For stars ${<}0.7$ \textit{SDSS} $r$-band magnitudes below MS turnoff. $\sigma$ is a 90\%-confidence upper bound. $N$ is from their Figure~13.\vspace{0.08cm}} \\[0.07cm]
GD-1       & --       & \parbox[m]{\linewidth}{\vspace{0.08cm}\raggedright\cite{MalhanIbata2019}\vspace{0.08cm}} & 67  & PM & ${<}2.3$ & ${\approx}$12 & \parbox[m]{\linewidth}{\vspace{0.08cm}For stars ${<}1.0$ \textit{Gaia} $G$-band magnitudes below MS turnoff. $\sigma$ is a 95\%-confidence upper bound. We estimate $v_{\rm cut}$ from their $5\sigma$ clipping of PMs.\vspace{0.08cm}} \\[0.07cm]
GD-1       & --       & \cite{Bonaca2020}      & 43  & RV & ${<}1$ & 7 & \parbox[m]{\linewidth}{\vspace{0.08cm}For stars ${<}2.1$ \textit{Pan-STARRS} $g$-band magnitudes below MS turnoff.\vspace{0.08cm}} \\[0.07cm]
GD-1       & --       & \cite{Malhan2022b}     & --  & PM & $0.63^{+0.67}_{-0.45}$ & NA &  \\[0.07cm]
GD-1       & --       & \cite{Valluri2025}     & \makecell{${\approx}95$\\(58)} & RV & \makecell{$2.85\pm0.6$\\($4.84\pm1.13$)} & 30 & \parbox[m]{\linewidth}{\vspace{0.08cm}For stars ${<}2.8$ \textit{DECaLS} $r$-band magnitudes below MS turnoff. The 1st (2nd) value is for the stream's cold (warm) component. $N$ is from their Figure~10.\vspace{0.08cm}} \\[0.07cm] \hline
Ophiuchus  & --       & \cite{Caldwell2020}    & 18  & RV & $1.6\pm0.3$ & NA & \parbox[m]{\linewidth}{\vspace{0.08cm}Includes only giant stars.\vspace{0.08cm}}\\[0.07cm]
Ophiuchus  & --       & \cite{Li2022}          & 118 & RV & $2.4 \pm 0.3$ & -- & \parbox[m]{\linewidth}{\vspace{0.08cm}$v_{\rm cut}$ used but unspecified.\vspace{0.08cm}}\\[0.07cm] \hline
300S       & --       & \cite{Li2022}          & 56  & RV & $2.5^{+0.4}_{-0.3}$ & -- & \parbox[m]{\linewidth}{\vspace{0.08cm}$v_{\rm cut}$ used but unspecified.\vspace{0.08cm}}\\[0.07cm] \hline
AAU        & --       & \cite{Li2022}          & 85  & RV & $4.3 \pm 0.4$ & 25 & \parbox[m]{\linewidth}{\vspace{0.08cm}$v_{\rm cut}$ from companion paper \citep{Li2021}.\vspace{0.08cm}}\\[0.07cm] \hline
Jet        & --       & \cite{Li2022}          & 32  & RV & $0.7^{+0.4}_{-0.5}$ & -- & \parbox[m]{\linewidth}{\vspace{0.08cm}$v_{\rm cut}$ used but unspecified.\vspace{0.08cm}}\\[0.07cm] \hline
Phoenix    & --       & \cite{Li2022}          & 26  & RV & $2.5 \pm 0.7$ & 9  & \parbox[m]{\linewidth}{\vspace{0.08cm}$v_{\rm cut}$ from companion paper \citep{Wan2020}.\vspace{0.08cm}}\\[0.07cm] \hline
Wilka Yaku & --       & \cite{Li2022}          & 9   & RV & $0.4^{+0.8}_{-0.4}$ & -- & \parbox[m]{\linewidth}{\vspace{0.08cm}$v_{\rm cut}$ used but unspecified.\vspace{0.08cm}}\\[0.07cm]
\enddata
\vspace{-0.04cm}\tablecomments{\footnotesize  Stream name, progenitor cluster (if known), reference, total members $N$ used to compute $\sigma$, velocity type (RV = line-of-sight radial velocities, PM = proper motions), $\sigma$, and the velocity cut $v_{\rm cut}$ used (if any) to exclude outliers. The last column comments on observational limits on membership or uncertainties. Uncertainties on $\sigma$ otherwise indicate a 16\%--84\% confidence interval. Entries of a dash indicate streams with unknown progenitors, or that the data was unspecified in the reference. `NA' indicates when no explicit $v_{\rm cut}$ was used; instead velocity outliers were down-weighted when selecting stream members.}
\end{deluxetable*} 

\subsection{Comparison to Observed Stream Velocity Dispersions} \label{S:discussion_observed_dispersions}
As noted in Section~\ref{S:intro}, many observed streams determined to have originated from GCs have low $\sigma$; see Table~\ref{table:observed_vdisps}. With few exceptions, these streams have $\sigma\lesssim 3\,\kms$, typically higher than we predict from internal GC dynamics alone, but not as high as predicted for GCs originating in cuspy DM halos or subject to heating by cold DM subhalos \citep[e.g.,][]{Malhan2021a, Malhan2022b, Carlberg2024b, Carlberg2025}. Further accounting for the other factors above that also enhance observed $\sigma$, including potentially an $m$--$\sigma$ correlation in some streams, would only exacerbate this mild tension between observed $\sigma$ and predictions for cold DM.

Membership selection for many GC streams is limited to stars not much below the main-sequence (MS) turnoff---see Table~\ref{table:observed_vdisps}'s last column---i.e., to stars of above-average mass. Such streams are potentially susceptible to the $m$--$\sigma$ correlation we find at such masses in streams from core-collapsed GCs. However, none of the six tabulated streams with a specific known GC progenitor are core-collapsed \citep{Trager1995}, suggesting negligible $m$ dependence in their measured $\sigma$. The other tabulated streams have no known progenitor, but the $m$--$\sigma$ correlation could affect the inferred $\sigma$ of some since ${\approx}20\%$ of GCs in the MW are core-collapsed \citep{Trager1995}, including two with stellar streams: M30 and NGC~6397 \citep{Bonaca2021}. Other GCs with streams, such as M2, M3, M92, NGC~1851, and NGC~5824 have borderline cuspy surface brightness profiles and are highly mass-segregated \citep{Weatherford2020}, indicating low BH retention and significant progression towards core collapse. So the streams from these GCs may be more likely to exhibit an $m$--$\sigma$ correlation from strong encounters.

Many streams also likely originate from fully dissolved GCs, but it is hard to predict what fraction of these underwent core collapse. Whether the GC's lifetime increases or decreases with cluster density depends on how \textit{tidally filling} the GC is---i.e., on the ratio of the GC's half-mass radius to its tidal radius. Central heating from BH binaries cause diffuse, low-$N$ GCs to overflow their tidal boundary, dissolving faster with lower density \citep{Gieles2021, Gieles2023, Roberts2024}. In this case, the GCs retain robust BH populations until dissolution, and never undergo core collapse. But GCs of roughly typical density in the MW or higher, especially those subject to weak tides (i.e., at high Galactocentric distances), dissolve faster with higher density due to their faster relaxation \citep[e.g.,][]{Baumgardt2001, BinneyTremaine2008}; see also some additional discussion in Section~6.3 of \cite{Weatherford2024}. In this case, the GCs can eject their BHs well before dissolution, giving rise to the observed core-collapsed GCs in the MW. But the large uncertainties on the initial properties of dissolved GCs, and the several factors at play (masses, central densities, tidal radii), make it challenging to predict the fraction of dissolved GCs that underwent core collapse and potentially produced streams with robust $m$--$\sigma$ correlations. So for now we simply note the low $\sigma$ of many GC streams, with membership limited to near the MS turnoff or above, makes strong encounter ejecta potentially relevant to DM inference based on $\sigma$.

\subsection{Fluctuating Tides and Direct $N$-body Comparison} \label{S:discussion_nbody}
By assuming the GC circularly orbits the Galactic center in a smooth\revision{, static} underlying potential, our mock streams do not account for fluctuating tides from passage near MW substructure, e.g., the Galactic center \revision{or evolution of the Galactic potential. These effects would} primarily enhance stripping of the GC's outermost stars, preferentially adding low-energy escapers to the stream and thereby reducing the fraction of stream members ejected by strong encounters \citep{Panithanpaisal2025}. This could conceivably wash out the $m$--$\sigma$ correlation from dynamics in the GC's core. But due to mass segregation, fluctuating tides mostly strip a GC's lowest-$m$ stars. So fluctuating tides may have little influence on $\sigma$ at $m$ above average in a GC (${\approx}0.5\,\Msun$, i.e., readily observable $m$), preserving the $m$--$\sigma$ correlation. This is best explored with direct $N$-body models due to their more rigorous accounting for Galactic tides. We thus examined kinematics in a few snapshots of streams from direct $N$-body models of the nearest core-collapsed GC, NGC~6397, that used the GC's true eccentric orbit---kindly provided by \cite{ArnoldBaumgardt2025}. Yet modeling differences made the comparison inconclusive, so we only describe it briefly and leave more careful testing to future work.

In summary, both the provided direct $N$-body snapshots, and some from \texttt{CMC} \citep{Rui2021a}, exquisitely match NGC~6397's observed surface brightness and velocity dispersion profiles (SBPs and VDPs). This inspired us to check if $\sigma$ appears similar in each set of stream models. Promisingly, they both exhibit similar $\sigma\approx1.5\,\kms$ for $m\lesssim0.5\,\Msun$, hinting that fluctuating tides from an eccentric GC orbit do not dramatically alter stream kinematics.

Unlike our models, the direct $N$-body snapshots did not exhibit an $m$--$\sigma$ correlation for $m>0.5\,\Msun$. This does not necessarily contradict our results, however. Due to $\sigma$'s noisiness (Figure~\ref{fig:vdisp_indiv_snaps_tot}), individual snapshots often show no correlation by pure chance. The discrepancy is also easily attributable to modeling differences. To hasten computations, the direct $N$-body models lacked primordial binaries and used different initial conditions tuned to result in a match to NGC~6397's SBP and VDP after only $4\,\Gyr$ versus NGC~6397's true age, $13\,\Gyr$ (also the rough age of the well-fitting \texttt{CMC} snapshots). For example, they began with a stellar mass function similar to that observed in NGC~6397 today, much flatter than the standard \cite{Kroupa2001} initial mass function in \texttt{CMC}. As expected for such differences, the well-fitting direct $N$-body snapshots were less mass-segregated than those from \texttt{CMC}. This, and the lack of primordial binaries\footnote{3BBF yields some binaries but is not sufficient to fully replace primordial binaries \citep{HeggieHut2003}. Significant 3BBF rates also require robust BH populations---reasonably excluded in the direct $N$-body models since these were initialized much closer to this BH-depleted GC's present.} to drive strong ejections, means there is no reason to expect an $m$--$\sigma$ correlation in the direct $N$-body snapshots anyway. We intend to conduct testing on more even footing in the future.

\subsection{Unresolved Binaries} \label{S:discussion_binaries}
Although we excluded binaries from our analysis in Section~\ref{S:results}, we share a few details about them here due to interest in the properties of unresolved binaries in streams and their impact on $\sigma$ from radial velocity measurements \citep[e.g.,][]{Li2019b}. The fraction of stars in our streams with binary companions is ${\approx}4.5\%$, dropping to ${\approx}3.5\%$ if limiting to primary star masses $\gtrsim0.7\,\Msun$, due to mass segregation and preferential escape of low-mass stars. These fractions depend negligibly on GC age or $r_v$, likely since \texttt{CMC} only fully considers hard binaries \citep{Heggie1975}, which are not easily disrupted by increases in GC density. They also change negligibly with initial $f_{b,h}$ (due to rarity of stars with $m>15\,\Msun$), or if we consider only binaries with $v \leq 20\,\kms$. These values are not especially predictive, though, as they mainly reflect \texttt{CMC}'s default $5\%$ initial hard binary fraction, albeit reasonable for MWGCs of typical density \citep{Milone2012}. 

Regarding the impact on $\sigma$ from unresolved binaries, note that those with internal orbital speeds $v_{\rm orb}\gg\sigma$ are often easily filtered out as outliers during stream membership selection (depending on inclination to the line-of-sight). Even for $r_v=2\,\pc$, 85\% of our escaped binaries have semi-major axes ${\lesssim}0.3\,{\rm au}$ \citep[Figure~6 of][]{Weatherford2023}, implying $v_{\rm orb}{\gtrsim}50\,\kms$ and easy filtration, especially if informed by model stream velocity distributions such as presented here. Wider binaries with lower $v_{\rm orb}$ pose a larger threat of slipping through the filter and artificially inflating $\sigma$. Both the destruction and formation rates of wide binaries in GCs are very high, with pairs of stars constantly being perturbed into and out of bound states, so there are always some wide binaries present despite their typically short lifespans \citep{Heggie1975, GoodmanHut1993}. While survival long enough to enter a stream is most likely in GCs of lower mass or density than considered here, or during the GC's final dissolution phase \cite[e.g.,][]{Kouwenhoven2010, MoeckelClarke2011}, the impact on $\sigma$ certainly merits future investigation.

A second interesting trend in our streams at ages typical of MWGCs is that the fraction of stars with a compact object companion increases roughly exponentially with $m$ over the range $m\in[0.1,0.8]\,\Msun$. The fraction rises from ${\approx}0.003\%$--$0.8\%$ ($r_v=0.5\,\pc$) or from ${\approx}0.01\%$--$0.3\%$ ($r_v=2\,\pc$). The trend with $m$ reflects the preference of resonant BS/BB encounters to pair the two highest-$m$ bodies in the interaction \citep{Heggie1975}. The steeper trend with lower $r_v$ reflects the higher rate of BS/BB encounters, giving more chances to pair massive bodies together. We direct readers interested in more detailed properties of our GC's binary ejecta (e.g., star types, semi-major axes, eccentricites, and mass ratios) to Section~4.4 and Figures~5--6 of \cite{Weatherford2023}.

\section{Summary and Future Work} \label{S:summary}
We explored the impact of internal dynamics in GCs on the kinematics of their stellar streams using Monte Carlo $N$-body simulations under a static Galactic tide. Full simulation of strong encounters, mass segregation, and BH populations in the GC's core enabled us to generate 
mock streams with realistic internal mass and velocity distributions. Our simulation initial conditions, specifically their initial $r_v=(0.5,2)\,\pc$, produce clusters representative of typical core-collapsed and non-core-collapsed MWGCs. We analyzed the kinematics of stream stars from each of these archetypes based on several velocity metrics: in Section~\ref{S:results}, the stars' absolute velocities relative to the GC's orbital velocity in Galactocentric cylindrical coordinates---$v=\lvert \vec{v_\star}-(0,v_c,0)\rvert$ with dispersion $\sigma_v$---and in the Appendix, the absolute value of each of $v$'s components: ($v_R,v_\phi,v_Z)$ with dispersions $(\sigma_R,\sigma_\phi,\sigma_Z)$.

In the GCs born less dense ($r_v=2\,\pc$), prolonged retention of a central BH population and corresponding dynamical heating prevents the onset of core collapse by the $9$--$13\,\Gyr$ age range typical of MWGCs, quenching mass segregation among luminous stars \citep{Alessandrini2016, Peuten2016, Webb2017, Weatherford2018, Weatherford2020}. In the streams from these GCs, $\sigma$ increases somewhat with star mass $m\in[0.1,0.8]\,\Msun$, rising from $\sigma_v\approx1.2$--$2.6\,\kms$---or $1.2$--$1.6\,\kms$ when counting only stars with $(v_R,v_\phi,v_Z)\leq20\,\kms$, mimicking kinematic membership cuts in stream observations. But faster loss of BHs in the denser GCs ($r_v=0.5\,\pc$) allows the core to collapse by $9$--$13\,\Gyr$ and the highest-$m$ stars to segregate deeper into the core. The resulting stronger correlation between $m$ and likelihood of ejection by strong encounters yields a steeper $m$--$\sigma$ correlation after core collapse, increasing from $\sigma_v\approx1.2$--$4.0\,\kms$, or $1.2$--$2.2\,\kms$ when limiting to stars with $(v_R,v_\phi,v_Z)\leq20\,\kms$. The ejecta from strong encounters also produce robust exponential tails in the stream's velocity distribution, qualitatively similar to the contribution of warm cocoons surrounding streams like GD-1.

In our streams, $\sigma$ at high $m$ fluctuates greatly depending on when the stream is ``observed" and which specific ejecta are identified as members. This effect is a result of small sample size in the number of high-$m$ ejecta from strong encounters rather than the number of high-$m$ ejecta in the stream overall. The 5th--95th percentile range of these fluctuations span $\sigma_v\in[1.3,7.3]\,\kms$ for stars with $m\geq 0.7\,\Msun$ after core collapse in the denser GCs, but reduce to $\sigma_v\in[1.1,3.5]\,\kms$ when counting only stars with $(v_R,v_\phi,v_Z)\leq20\,\kms$. It is thus possible that observations of streams from GCs limited to stars near and above the MS turnoff may sometimes measure $\sigma\gtrsim3\,\kms$, even in the absence of any heating from DM subhalos or other MW substructure, any preprocessing in a parent halo, or any spurious enhancement from, e.g., unresolved binaries. This large scatter also implies the overall $\sigma$ measured in any \textit{individual} stream is a somewhat imprecise measure of heating applied to that stream by MW substructure, or of properties of the stream's progenitor cluster or parent halo, though $\sigma$ can still rule out causes of especially large $\sigma$, e.g., origins in massive cuspy parent halos. The overall $\sigma$'s sensitivity to high-$v$ outliers also motivates future studies to report full velocity distributions---either in the form of raw data or parameters of mixture models fitted to the data, including the distribution's tail. Indeed, the high-$v$ tail is the portion of the velocity distribution most useful for DM inference \citep[e.g.,][]{Carlberg2024b, Carlberg2025}.

Our findings imply observations of streams from dense, highly mass-segregated GCs---e.g., those that are BH-depleted and core-collapsed---may be biased to somewhat higher $\sigma$ than would be predicted by typical stream models neglecting $\sigma$'s $m$-dependence. Accounting for such a bias would mildly reduce observationally inferred $\sigma$ in streams---perhaps ${\approx}1/3$ of GC streams if restricting to changes in $\sigma$ of ${\gtrsim}0.5\,\kms$ when using typical kinematic membership cuts. This would exacerbate tension between the already low $\sigma$ in many streams and expectations of significant heating by cold DM subhalos, especially in light of further enhancements to $\sigma$ expected from stream formation in a parent halo prior to accretion by the MW, heating from many baryonic MW substructures, and artificially enhanced $\sigma$ from unresolved binaries, contamination by background stars, or the dependence of gravitational redshift on star mass and radius. An evolving MW potential \citep[e.g.,][]{Panithanpaisal2025} can also further exacerbate any tail in the velocity distribution from strong encounters. An $m$--$\sigma$ correlation in streams would thus cause stream observations to further favor alternatives to cold DM, such as \revision{warm DM, observationally viable models of ULDM, and perhaps some SIDM models that significantly suppress the low-mass end of the subhalo mass function.}

An important caveat to our conclusions is that we limited our analysis to GCs on circular orbits in \revision{a smooth, static MW potential. Interactions with baryonic and DM substructure in a more realistic MW potential would enhance stream $\sigma$ regardless of $m$, so should leave the the $m$--$\sigma$ correlation intact. However,} GCs subject to strongly fluctuating tides \revision{(e.g., from eccentric GC orbits or rapid evolution in the MW potential) would exhibit} enhanced tidal stripping from the GC's halo, which could conceivably wash out the $m$--$\sigma$ correlation from dynamics in the GC's core. But since the tidal stripping would primarily affect the GC's lowest-$m$ stars, fluctuating tides may have little influence on $\sigma$ at $m$ above average in a GC (${\approx}0.5\,\Msun$, i.e., readily observable $m$), preserving the $m$--$\sigma$ correlation. We intend to explore such cases in future work, further examine the impact of unresolved binaries, and search for an $m$--$\sigma$ relation in observed streams. Regarding the latter, we have performed an initial search, finding suggestive evidence for an $m$--$\sigma$ correlation in the nearest GC stream (from the core-collapsed GC NGC~6397). But we defer concrete analysis to future work due to a desire for higher sample purity and concern that the correlation also appeared in the stream's background.

While we do not expect different choices of a static MW potential to effect the results, as it adds no dynamics capable of reversing or washing out the m-sigma correlation, it is possible that fluctuating tides originating from an evolving MW potential (or an eccentric cluster orbit) could sufficiently enhance tidal stripping of low-energy ejecta from the star cluster periphery to wash out/reduce the strength of our m–sigma relation (i.e., by reducing the fraction of strong encounter ejecta in the stream).

\vspace{-0.5cm}
\begin{acknowledgements}
We thank \revision{the anonymous referee,} Sergey Koposov, and Holger Baumgardt for valuable feedback that improved this work. S.K. suggested reporting the velocity distribution fits and each velocity component. H.B. provided data from direct $N$-body models for comparison.
\end{acknowledgements}

\begin{contribution}
N.C.W. conceived the project and led the data analysis and paper write-up. A.B. oversaw project progress, providing frequent input on content and interpretation.
\end{contribution}

\software{\texttt{CMC} \citep{CMC_v1.0}, \texttt{COSMIC} \citep{COSMIC}, \texttt{Gala} \citep{Gala_v1.3}, \texttt{Astropy} \citep{Astropy, Astropy2, Astropy3}, \texttt{matplotlib} \citep{Matplotlib}, \texttt{SciPy} \citep{SciPy}, \texttt{NumPy} \citep{NumPy}, \texttt{pandas} \citep{Pandas}.}

\appendix
\setcounter{figure}{0}
\renewcommand{\thefigure}{A\arabic{figure}}
\setcounter{table}{0}
\renewcommand{\thetable}{A\arabic{table}}
\setcounter{equation}{0}
\renewcommand{\theequation}{A\arabic{equation}}

\begin{figure*}[ht!]
\centering
\includegraphics[width=0.98\linewidth]{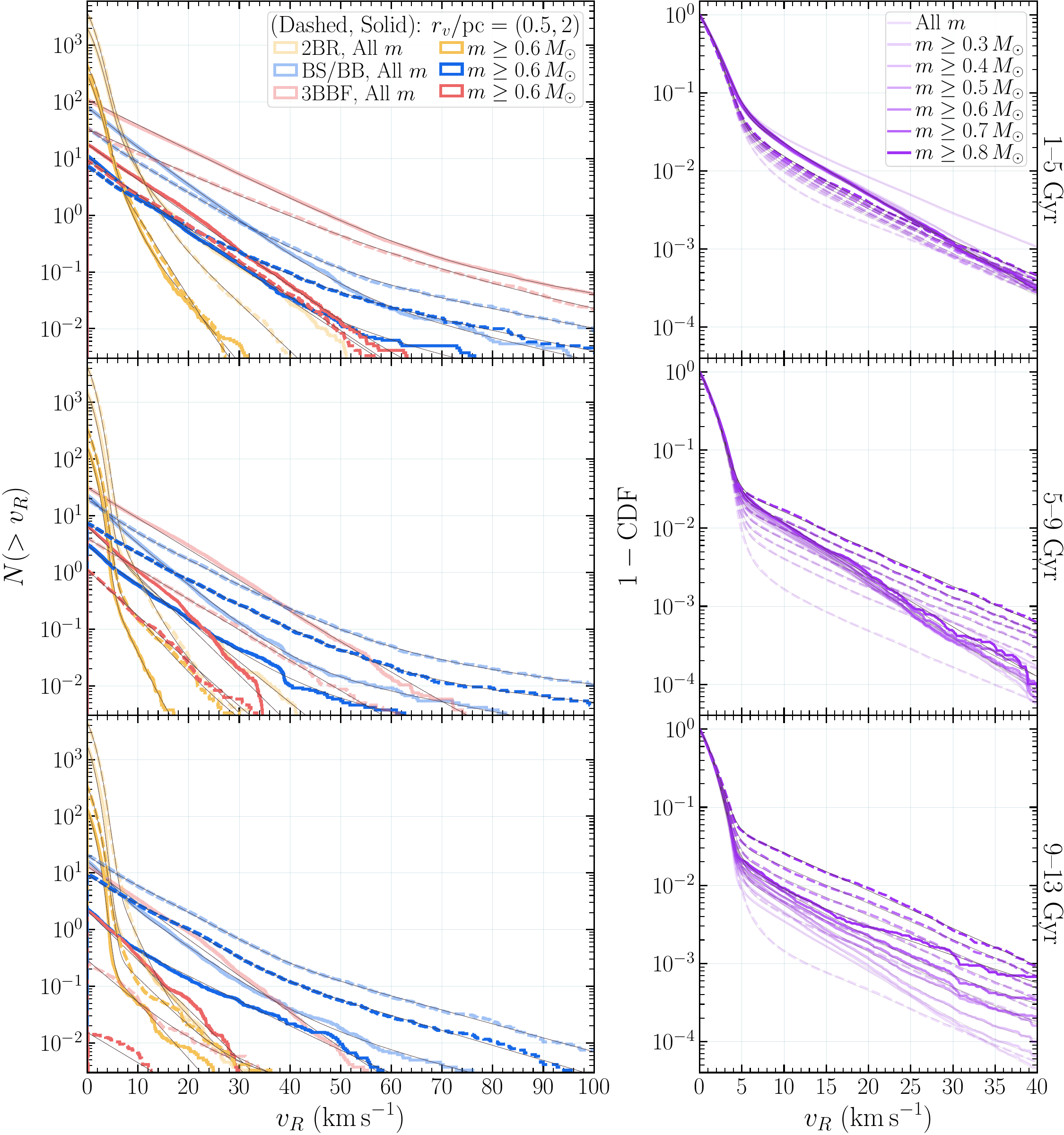}
\caption{Same as Figure~\ref{fig:vdist_tot}, except now for $v_R$.}
\label{fig:vdist_rad}
\end{figure*}

\begin{figure*}[p!]
\centering
\includegraphics[width=0.959\linewidth]{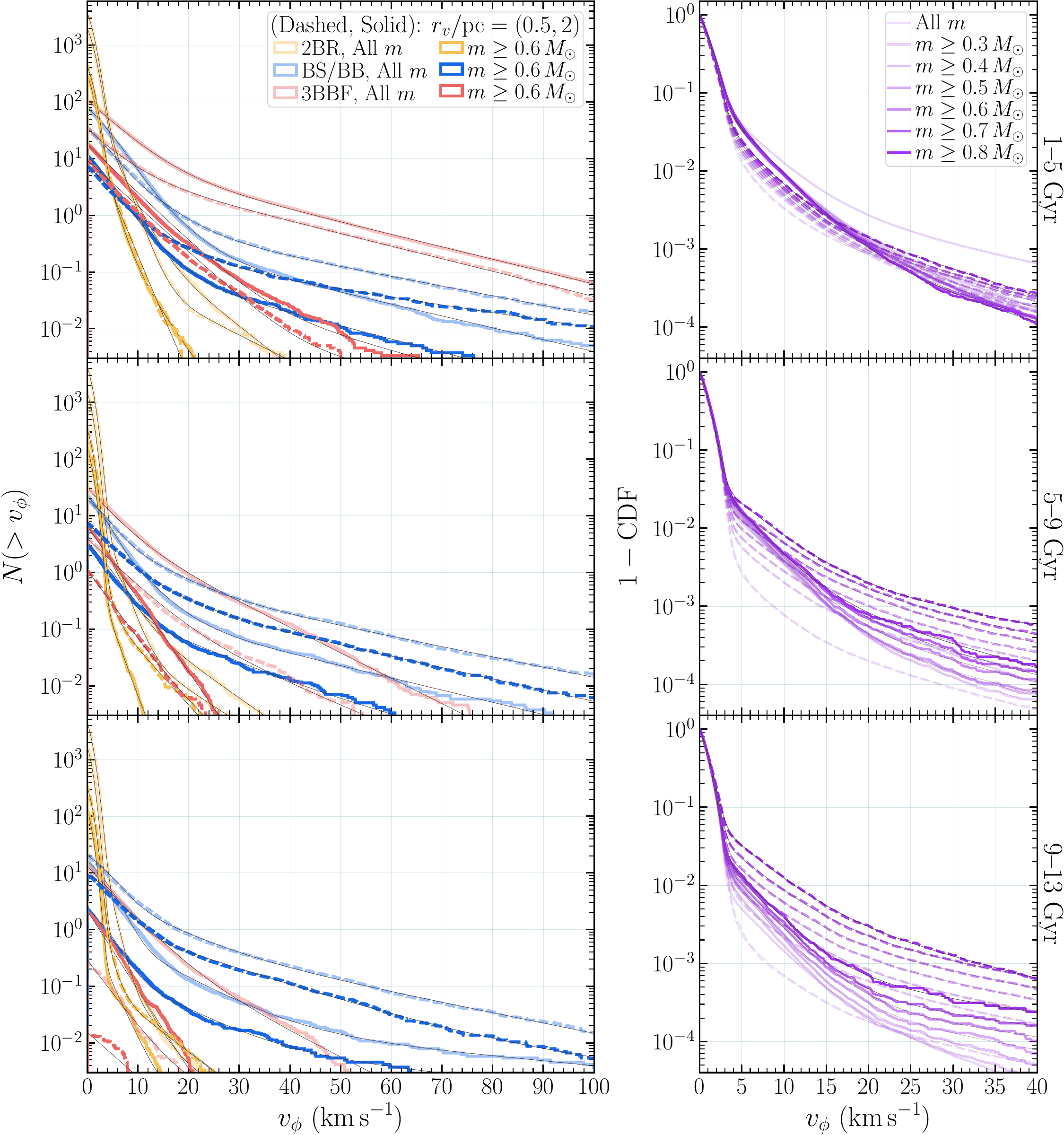}
\caption{Same as Figure~\ref{fig:vdist_tot}, except now for $v_\phi$.}
\label{fig:vdist_phi}
\end{figure*}

\begin{figure*}[p!]
\centering
\includegraphics[width=0.959\linewidth]{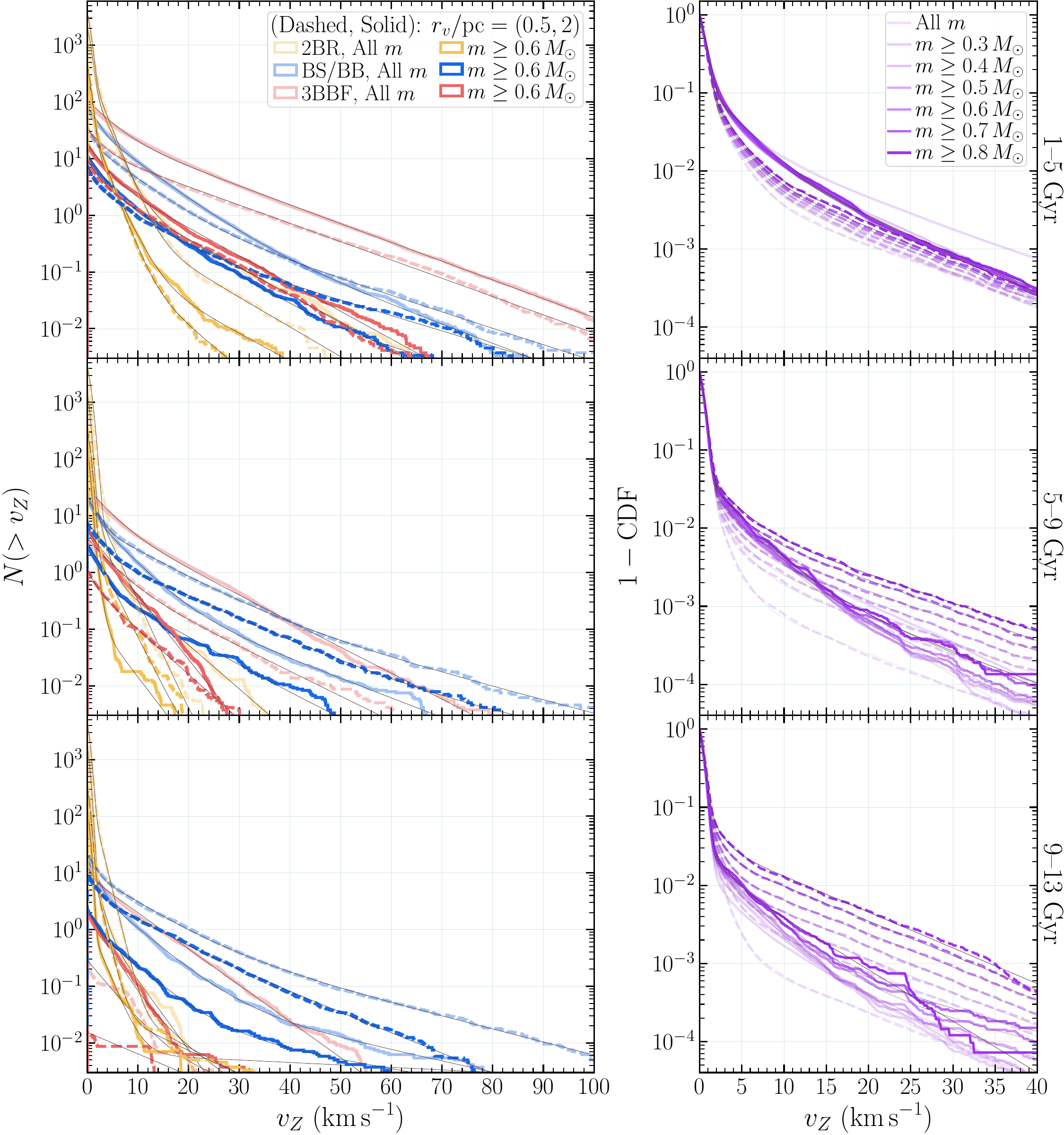}
\caption{Same as Figure~\ref{fig:vdist_tot}, except now for $v_Z$.}
\label{fig:vdist_z}
\end{figure*}

\begin{figure*}[ht!]
\centering
\includegraphics[width=0.88\linewidth]{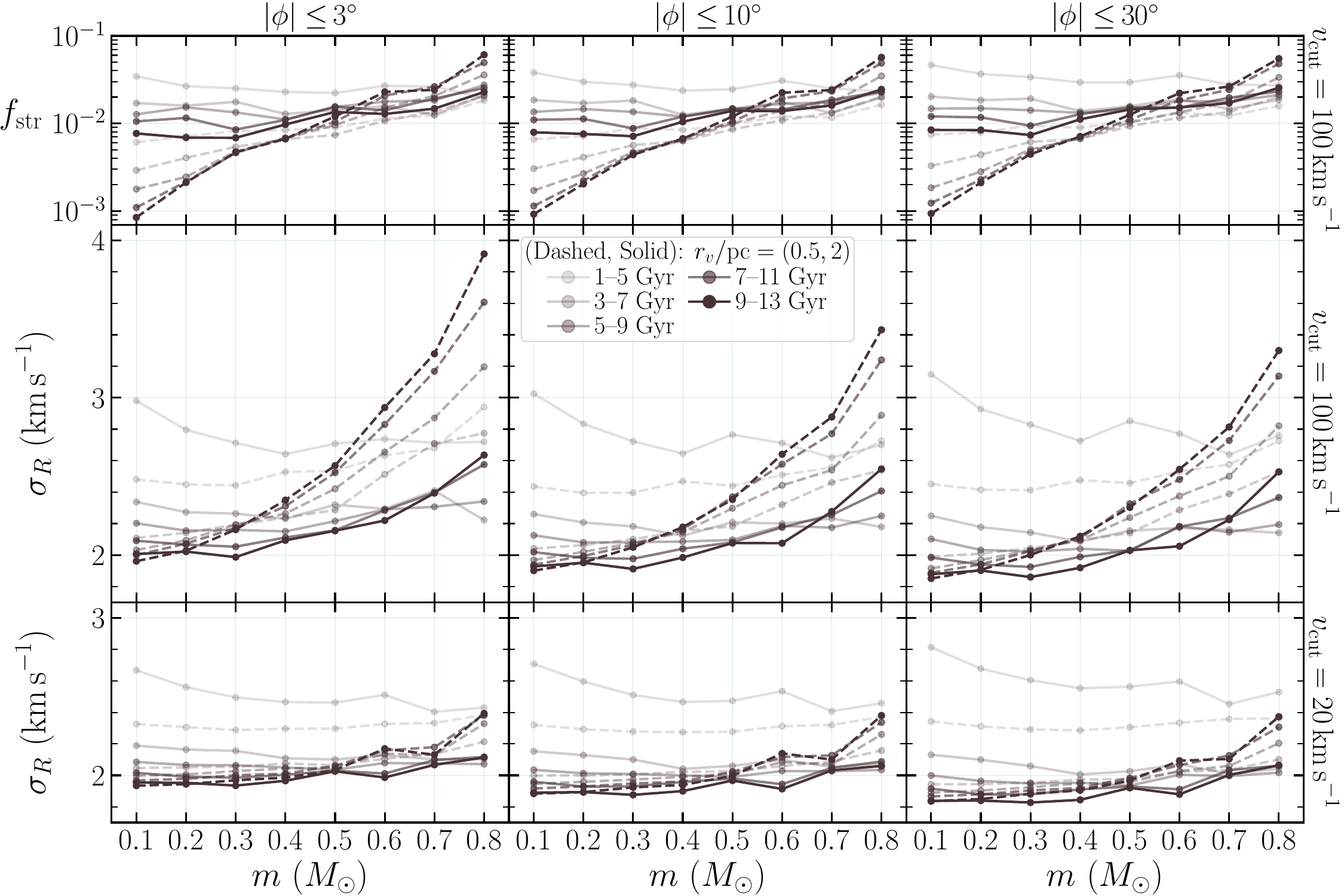}
\caption{Same as Figure~\ref{fig:vdisp_all_snaps_tot}, except now for $\sigma_R$.}
\label{fig:vdisp_all_snaps_rad}
\end{figure*}

\begin{figure*}[ht!]
\centering
\includegraphics[width=0.88\linewidth]{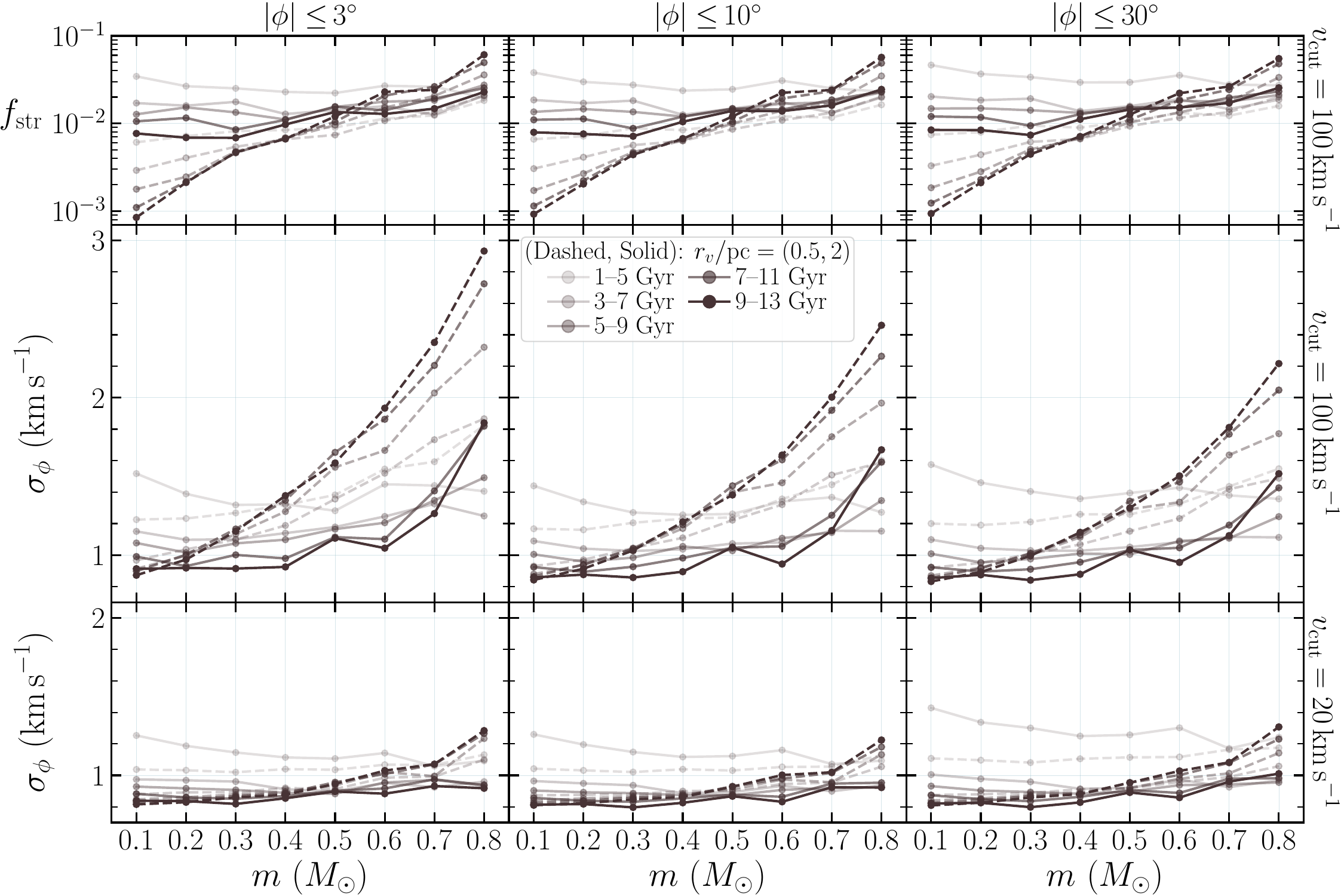}
\caption{Same as Figure~\ref{fig:vdisp_all_snaps_tot}, except now for $\sigma_\phi$.}
\label{fig:vdisp_all_snaps_phi}
\end{figure*}

\begin{figure*}[ht!]
\centering
\includegraphics[width=0.88\linewidth]{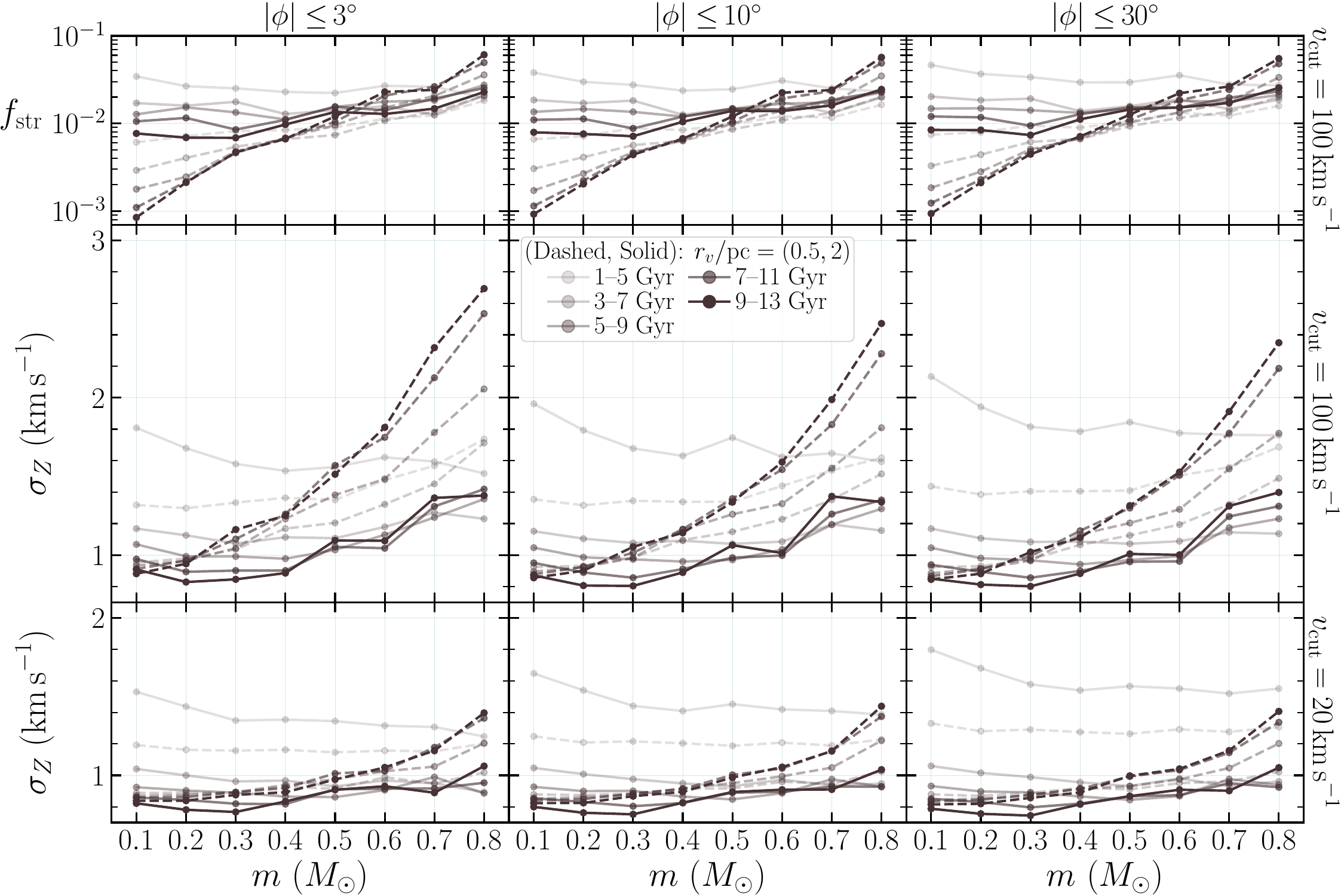}
\caption{Same as Figure~\ref{fig:vdisp_all_snaps_tot}, except now for $\sigma_Z$.}
\label{fig:vdisp_all_snaps_z}
\end{figure*}

\begin{figure*}[ht!]
\centering
\includegraphics[width=\linewidth]{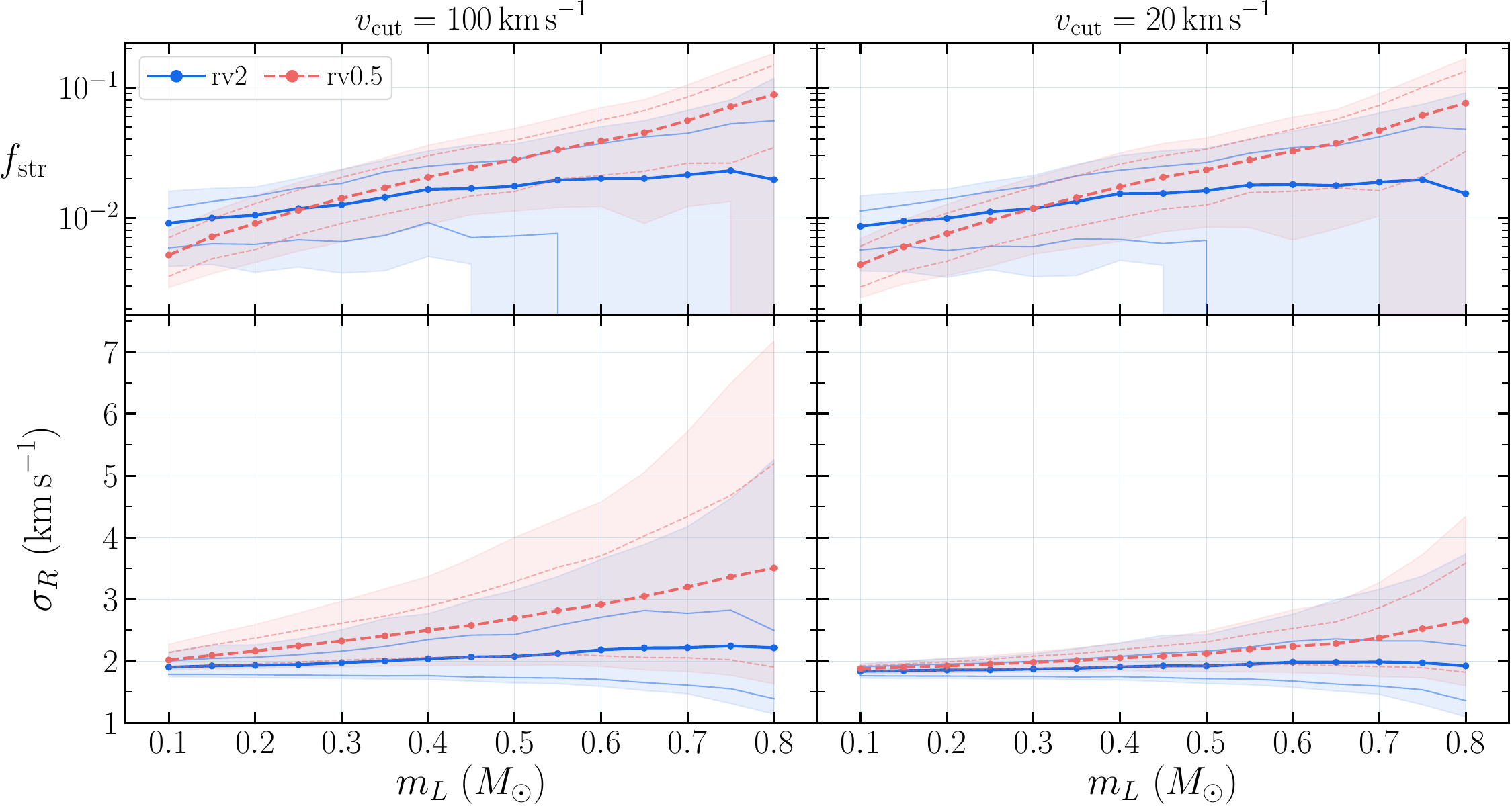}
\caption{Same as Figure~\ref{fig:vdisp_indiv_snaps_tot}, except now for $\sigma_R$.}
\label{fig:vdisp_indiv_snaps_rad}
\end{figure*}

\begin{figure*}[ht!]
\centering
\includegraphics[width=\linewidth]{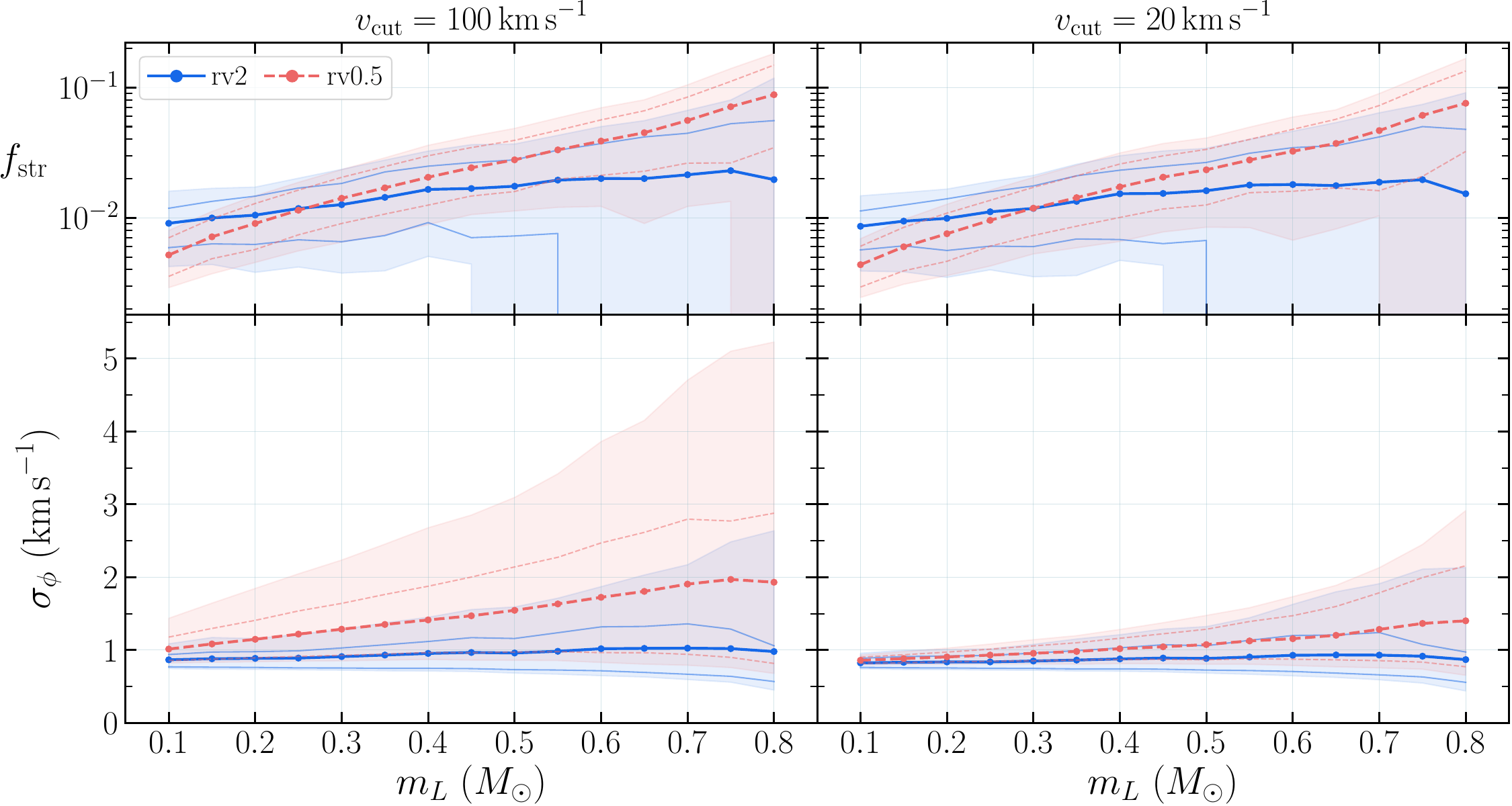}
\caption{Same as Figure~\ref{fig:vdisp_indiv_snaps_tot}, except now for $\sigma_\phi$.}
\label{fig:vdisp_indiv_snaps_phi}
\end{figure*}

\begin{figure*}[ht!]
\centering
\includegraphics[width=\linewidth]{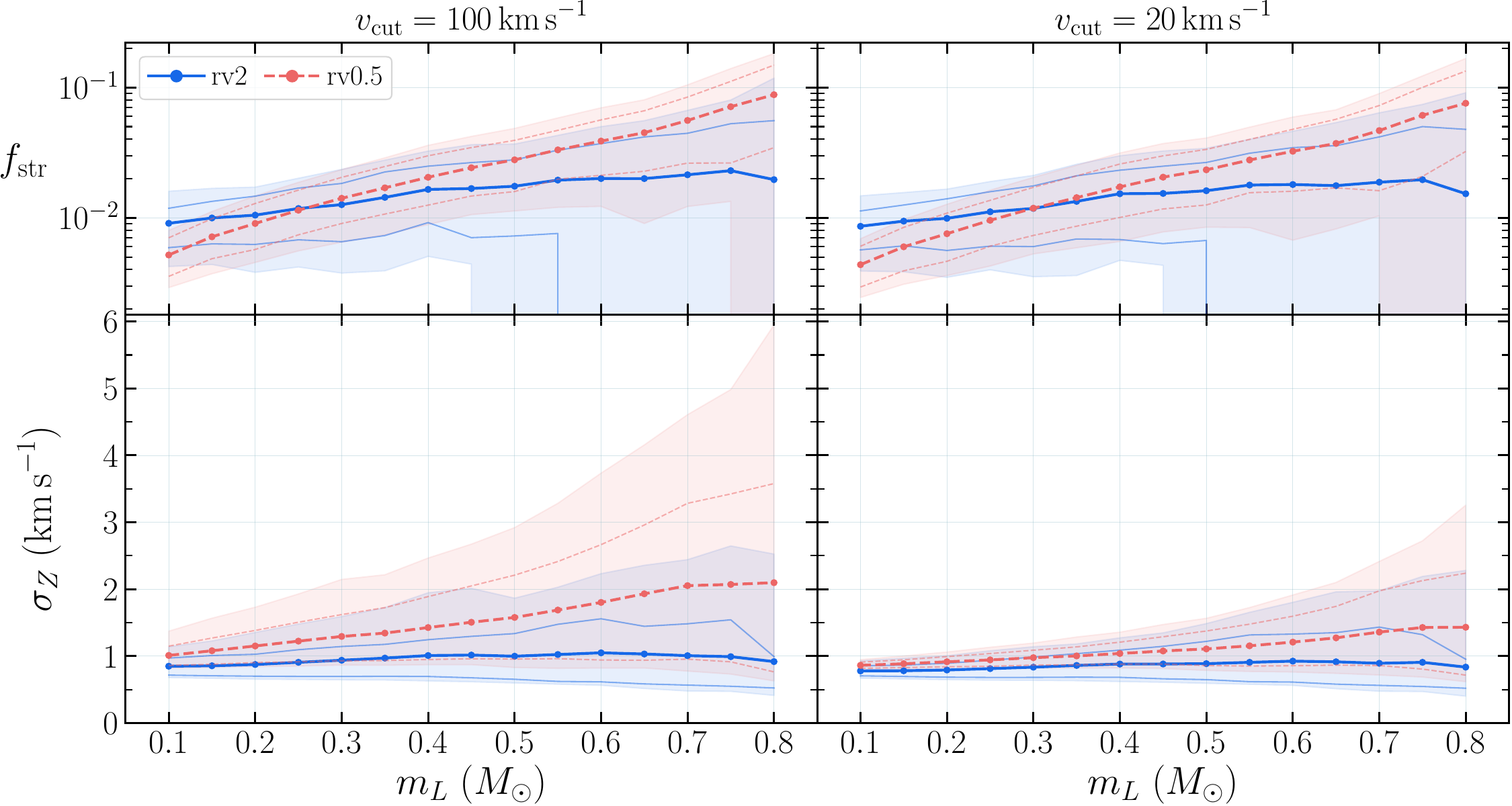}
\caption{Same as Figure~\ref{fig:vdisp_indiv_snaps_tot}, except now for $\sigma_Z$.}
\label{fig:vdisp_indiv_snaps_z}
\end{figure*}

We here report our streams' kinematics in terms of the absolute values of $v$'s component velocities in Galactocentric cylindrical coordinates $(v_R,v_\phi,v_Z)$, and their corresponding dispersions $(\sigma_R,\sigma_\phi,\sigma_Z)$. We show versions of Figures~\ref{fig:vdist_tot}--\ref{fig:vdisp_indiv_snaps_tot} for each such component $v_i$ (and corresponding $\sigma_i$) in Figures~\ref{fig:vdist_rad}--\ref{fig:vdisp_indiv_snaps_z}. The parameters of the fits to the component $v_i$ distributions are accessible in the full machine-readable version of Table~\ref{table:curve_fits_rel} in the online journal, though note that the Maxwellian terms in the fits for $v$ overall, and the individual contributions from 2BR and strong BS/BB encounters, are now half-normal distributions (half since we define each $v_i\geq0$). The corresponding version of Equation~(\ref{eq:mee_fit}) from Section~\ref{S:results_curve_fitting} is thus
\begin{equation} \label{eq:nee_fit} \begin{aligned}
P_{1i}(v_i|&\sigma_1,\sigma_2,\sigma_3,a_1,a_2) =\\
&\frac{a_1}{\sigma_1} \sqrt{\frac{2}{\pi}} e^{-\frac{v_i^2}{2\sigma_1^2}} + \frac{a_2}{\sigma_2}e^{-\frac{v_i}{\sigma_2}} + \frac{(1-a_1-a_2)}{\sigma_3}e^{-\frac{v_i}{\sigma_3}}.
\end{aligned} \end{equation}
The corresponding version of Equation~(\ref{eq:ne_fit}) expressing the contribution to each component $v_i$ from 3BBF is instead, 
\begin{equation} \label{eq:ee_fit}
P_{2i}(v_i|\sigma_1\sigma_2,a_1) =\\
\frac{a_1}{\sigma_1} e^{-\frac{v_i}{\sigma_1}} + \frac{1-a_1}{\sigma_2} e^{-\frac{v_i}{\sigma_2}}.
\end{equation}

\clearpage
\bibliography{main}
\end{document}